\pdfoutput=1
\documentclass[preprint, nonatbib, 10pt]{sigplanconf}
\usepackage[numbers,sort&compress]{natbib}
\usepackage{amsmath, amssymb, amscd}
\usepackage{amsfonts}
\usepackage{graphicx}
\usepackage{float}
\usepackage{subfig}
\usepackage{flushend} 
\usepackage{multirow}
\usepackage{times}
\usepackage{listings}
\usepackage{verbatim}
\usepackage{color}
\usepackage{xcolor}
\usepackage{courier}
\usepackage{epsfig}
\usepackage{txfonts}
\usepackage{setspace}

\usepackage{url}
\usepackage[colorlinks=true,linkcolor=red,urlcolor=blue,citecolor=blue]{hyperref} 
\usepackage{breakurl}
\usepackage{placeins}
\usepackage{listing}
\usepackage[normalem]{ulem}     

\usepackage[maxfloats=150]{morefloats}
\usepackage{microtype} 
\usepackage[utf8]{inputenc}

\usepackage[iso,american,cleanlook]{isodate} 
\usepackage{oldlfont,epsfig}
\usepackage{epic}
\usepackage{eepic}

\usepackage{todonotes} 
\usepackage{rviewport} 
\usepackage{fix-cm} 

\newenvironment{ItemizeTight}{
\begin{itemize}
  \setlength{\itemsep}{2pt}
  \setlength{\parskip}{1pt}
  \setlength{\parsep}{1pt}}{\end{itemize}
}

\begin{document}

\setlength{\pdfpageheight}{\paperheight}                                             
\setlength{\pdfpagewidth}{\paperwidth}                                               
                                                                                     
%% \conferenceinfo{PPoPP 2017} {February 4--8, 2017, Austin, Texas, USA.}
%% \copyrightyear{2017}
%% \copyrightdata{978-N-NNN-NNNN-N/NN/NN}
%% \doi{nnnnnn.nnnnnn} 

%% \titlebanner{TitleBanner}
%% \preprintfooter{PPFooter} 
\title{Malthusian Locks} 
%% \subtitle{PPoPP 2017 submission \# 54} 
%% \subtitle{EuroSys 2017 submission \# 143} 

%% \titlebanner{...}
%% [DRAFT] \preprintfooter{draft}
%% \toappear{Under submission}

\authorinfo{\vspace{-10 pt}Dave Dice}
           {Oracle Labs}
           {dave.dice@oracle.com}

%% \date{\today}
\maketitle

%% \parindent 0pt

%% ===================================================================
%% Invisible text : 
%% Use to make terms searchable in pdf -- via spotlight or google -- 
%% but invisible on the page when printed or displayed with a pdf viewer.  
%% Enhanced SEO 
%%
%% Terms : hidden text; invisible text; 0-sized text; search-only; non-display; microdot
%%
%% See : Invisible.tex
%%
%% Beware: excessively small font sizes for "invisible" operator family
%% appear in PS but not in PDF!
%% See main.log : 
%% LaTeX Font Warning: Font shape `OT1/cmr/m/it' in size <0.01> not available
%% Corresponds to \fontsize{0.01}{0.01} ...
%% Fix and work-around : revert to larger fonts : \fontsize{0.5}{0.5} 
%% 0.5 seems to suffice
%%
%% ===================================================================

%% The best implementation seems to be via \marginpar

\newcommand\Invisible[1]{
  \marginpar{\color{white}{\fontsize{.5}{.5}\selectfont #1 }}
}

%% Consider:
%% *  linespread; setspace; setstrech; baselineskip; setspace; baselinestretch
%% *  \setlength{\marginparwidth}{16mm}
%%    \setlength{\marginparsep}{2mm} 
%%    \setlength{\marginparpush}{0pt}
%% *  \setmarginnotes{0.1\foremargin}{0.7\foremargin}{\onelineskip}
%% *  marginpar vs marginnote
%% *  partial overstrike of adjacent lines
%% *  marginpar with negative vspace tricks
%% *  Marginal; Marginalia
%% *  BEWARE: marginpar is a float, as is footnote
%%    Usage restriction
%%    floats can not appear within floats, otherwise we see :
%%    LaTeX Error: Float(s) lost.
%%    By implication, Invisible{} implemented with marginpar can not 
%%    appear nested within another Invisible{}
%%    Similarly, there can be no nesting of Invisible and footnote.  
%% *  pdfcomment

\newcommand\InvisibleInMargin[1]{ 
  %% \marginpar{\color{white}{\scalebox{0.01}{#1}}}
  %% \marginpar{\color{white}{\scalebox{0.01}{\parbox{\linewidth}{#1}}}}
  %% \marginpar{\color{black}{\scalebox{0.4}{\parbox{8\linewidth}{#1}}}}
  \marginpar{\color{black}{\scalebox{0.4}{\begin{minipage}{8\linewidth}{#1}\end{minipage}}}}
}

\newcommand{\Boxer}[1]{%
  \makebox[0pt][l]{%
    \makebox[\linewidth+\marginparsep][l]{}%
    \parbox[t][0pt][t]{1\marginparwidth}{#1}%
  }%
} 

\newcommand\InvisibleORIGINAL[1]{
  {\color{white}{\fontsize{.5}{.5}\selectfont \setlength{\parskip}{0pt} \setlength{\parfillskip}{0pt}  {#1}}}
}

\newcommand\InvisibleQuote[1]{
 \begingroup 
 \color{white}\fontsize{1}{1}\selectfont 
 \begin{quote} {#1} \end{quote} 
 \endgroup
}

%% IBox = Invisible Box; for inline text
\newcommand{\IBox}[1]{\mbox{\color{white}\fontsize{.5}{.5}\selectfont {#1}}}

%% Floating footnote without a number
\newcommand{\AtFoot}[1]{\let\thefootnote\relax\footnotetext{{#1}}}

\newcommand\blfootnote[1]{%
  \begingroup
  \renewcommand\thefootnote{}\footnote{#1}%
  \addtocounter{footnote}{-1}%
  \endgroup
}

%% avoid situation where texttt extends into margin
\newcommand\wtt[1]{%
  \hfil\penalty0\hfilneg\texttt{#1}%
}

\newcommand{\RedText}[1]{\textcolor{red}{[#1]}}

\newcommand\RedBlock[1]{
  \begingroup\color{red}{#1}\endgroup
}

\newcommand\RedZone[1]{
  \begingroup\color{red}{#1}\endgroup
}

\newcommand\RedListingGrouped[1]{
  \begingroup\color{red}\begin{lstlisting}{#1}\end{lstlisting}\endgroup
}

\newcommand\RedListingFramed[1]{
  \begin{frame}[fragile]
  \color{red}\begin{lstlisting} {#1} \end{lstlisting}
  \end{frame} 
}

\newcommand\RedListing[1]{
  color{red}\begin{lstlisting} #1 \end{lstlisting}
}

%% Consider: \defverbatim\lst

\newcommand\RedQuote[1]{
 \begingroup 
 \color{red} 
 \begin{quote} {#1} \end{quote} 
 \endgroup
}

%% Inclusion-exclusion control ...
%% Variations:
%% manage long-extended and short-conference versions via
%% @  a single file-directory and use of ExtendedVersion|LVersion|Extended 
%%    to differentiate text -- conditional inclusion
%% @  by forking and creating a distinct subdirectory for the short version.  
%%    This can result in skew-drift and require synchronization of the
%%    the two versions.
%% 
%% LVersion-ExtendedVersion-Extended are rather like C/C++ #ifdefs and
%% conditional compilation. 
%%
%% Using comment-based TAG or MARK or OPTIONAL to designate areas
%% to be excluded or made invisible for the short-conference version.  
%% use \Exclude or \Invisible to elide-redact-cull.  
%% See "Invisible Text" below

\newcommand{\Xomit}[1]{}
\newcommand{\ignore}[1]{}
\newcommand{\Elide}[1]{}
\newcommand{\Exclude}[1]{}
\newcommand{\Include}[1]{#1} 
\newcommand{\Extra}[1]{#1} 
\newcommand{\Redact}[1]{} 
\newcommand{\Cull}[1]{} 
\newcommand{\ExtendedVersion}[1]{#1} 
\newcommand{\SubTrim}[1]{} 
\newcommand{\LVersion}[1]{#1} 
\newcommand{\ExVersion}[1]{#1} 
\newcommand{\Extended}[1]{#1} 
\newcommand{\Optional}[1]{#1} 
\newcommand{\Decide}[1]{#1} 
\newcommand{\Conditional}[1]{#1} 
\newcommand{\Marked}[1]{#1} 
\newcommand{\Tagged}[1]{#1} 
\newcommand{\Undecided}[1]{#1} 
\newcommand{\Relegate}[1]{#1} 

%% ================================================================
%% Edit pass -- check the following tags and decide to
%% either include or exclude from the paper.  
%% keywords: editorial pass; redact; scan; assay; survey; cull; resect; 
%% 
%% Tags : Optional; RedZone; IBox; Verbatim; Redundant; Redact; cull; excise; 
%% Key  : Exclude; Invisible; 
%%
%% CHECKLIST: Include; Exclude; Optional; Invisible; Redundant; 
%%
%% Consider:
%% Transiently-temporarily map-bind Invisible and Exclude macros to RedZone
%% then print resultant file
%% Red areas act as inline checklist -- tagged or marked text
%% Substitute-replace-map-bind-associate-remap-redefine-override-usurp
%% Ideally we'd use verbatim-red to mark conditionally excluded text.  
%%
%% ================================================================

%% 
%% \renewcommand\Invisible[1]{\RedZone {#1}}
%% \renewcommand\Exclude[1]{\RedZone {#1}}

%% \\ vs \hfill \break vs \newline
%% AKA: Bullet; Bulleted; Item; Point; 
\newcommand\Bullet[1]{* #1\noindent\hfill\break} 
\newcommand\Bull{\hfill\break\noindent~*}

%% V1.7 Computer Society "diamond line" which follows index terms for nonconference papers
%% V1.8a full width diamond line for single column use
\newcommand\IEEEDiamondLine{\vrule depth 0pt height 0.5pt width 4cm\nobreak\hspace{7.5pt}\nobreak
\raisebox{-3.5pt}{\fontfamily{pzd}\fontencoding{U}\fontseries{m}\fontshape{n}
\fontsize{11}{12}\selectfont\char70}\nobreak
\hspace{7.5pt}\nobreak\vrule depth 0pt height 0.5pt width 4cm\relax}

%=========================================================================
%  Abstract
%=========================================================================

%% Use either texttt or textsf
\newcommand{\malloc}{{\footnotesize \texttt{malloc}}}

\begin{abstract}

Applications running in modern multithreaded environments  
are sometimes \emph{overthreaded}. 
%% In modern multithreaded environments it is common to find situations
%% where an application is \emph{over-threaded}
%% not uncommon vs common : litotes
%% OPTIONAL ...
\blfootnote{Robert Malthus \cite{malthusianism} argued for population control, 
cautioning that societies would collapse as increasing populations competed for 
resources.  His dire predictions did not come to pass as food production -- which 
had previously been stagnant -- improved to keep pace with population growth.}
\blfootnote{This is an extended version of a paper appearing in EuroSys 2017:
\url{http://dx.doi.org/10.1145/3064176.3064203}.  Additional details can be found in 
\cite{CR-Patent}.} 
%% Appears in ; has been observed in 
The excess threads do not improve performance, and in fact may act to degrade
performance via \emph{scalability collapse} 
\footnote{Increased concurrency resulting in decreased throughput
appears in other contexts. Brooks\cite{Brooks1975} observed that increasing the 
number of workers on a project could slow delivery.},
which can manifest even when there are fewer ready threads than available cores.  
Often, such software also has
highly contended locks.   We \Invisible{opportunistically} leverage the existence of such locks 
by modifying the lock admission policy so as to intentionally limit
the number of distinct threads circulating over the lock in a given period.   
%% Wording: by modifying; and modify ; 
%% Wording: such locks; those locks
%% We leverage the existence of such locks and modify the lock admission policy to limit ...
%% Consider: eliminate "so as" ...
%% Variously: in a given period; over short periods; over the short term
%% Leverage existence of such locks ...
%% We utilize such locks 
%% In circulation at a given moment
%% In circulation in a given period
%% In circulation vs circulating over
Specifically, if there are more threads circulating than are necessary to keep 
the lock saturated (continuously held), our approach will selectively cull and 
passivate some of those excess threads.
%% of the threads circulating over that contended lock. 
We borrow the concept of \emph{swapping}
from the field of memory management and \Exclude{intentionally} impose
\emph{concurrency restriction} (CR) if a lock suffers from contention. 
%% WORDS: oversubscribed; saturated; contended; suffers contention; 
The resultant admission order is unfair over the short term but we explicitly 
provide long-term fairness by periodically shifting threads between the set of 
passivated threads and those actively circulating.  
%% OPTIONAL ...
%% In the worst case CR does no harm, but often yields performance benefits.  
Our approach is palliative, but is often effective at avoiding or reducing scalability collapse, 
and in the worst case does no harm. 
Specifically, throughput is either unaffected or improved, and unfairness
is bounded, relative to common test-and-set locks which allow unbounded bypass
and starvation  
\footnote{\emph{Bypass} occurs when a thread $T$ acquires a lock but there exist
other waiting threads that arrived earlier than $T$.}.  
%% In addition to 
By reducing competition for shared resources, such as pipelines, processors
and caches, concurrency restriction may also reduce overall resource consumption
and improve the overall load carrying capacity of a system. 

%% OPTIONAL
%% Refute-rebut concerns from PPoPP 2016 reviews regarding "no harm" claim
%% CR admits class of applications and pathology ...
\Invisible{Possible counter-argument or counter-example -- example of applications
that might suffer from unfair CR admission.  Imagine a ``ragged barrier'' which
does not satisfy rendezvous conditions until all participating threads have 
completed 10 loop steps.  Each step acquires and releases a contended CR-based lock
within the loop.
The time to reach rendezvous may be longer with a CR-based locks than with strict FIFO locks. 
} 

\Invisible{Drafty Draft} 

\Invisible{We draw an analogy-metaphor between threads and members of the populace. 
We anthropomorphize threads.} 

\Invisible{
*  Mitigate; attenuate; palliative; palliate; 
*  Saturate; provision; fill; satisfy; pack  
*  Saturated = Continuously held; contended; full occupancy   
*  Oversubscribed; Oversaturated; saturated; contended; Multiprogrammed; 
*  no laxity or slack time
*  suffers from contention; is contended; always has waiting threads; 
*  under-saturate; over-saturate; under-provision; over-provision; 
*  Carrying capacity; offered load; 
*  Less is more; fewer is faster; SIF = slower is faster; pyrrhic concurrency;  
*  Wasted or Pyrrhic Parallelism 
*  Contended vs fully saturated
*  Starvation
*  Denied entry; denied admission
*  primum non nocere; benign; 
*  palliative; avoidance; mitigation; remedy; provide relief; 
*  MRAT = Most-recently arrived thread
*  Require saturation and contention and waiting threads for a lock to 
   be able to decide which threads will be admitted.  Require surplus.  
*  sideline; passivate; arrest; detain; sequester; deactivate; suspend; capture;
*  Lock lore; folk myth; received wisdom; practicum; praxis; 
}

\Invisible{
*  CR   = MCSCRA8U; LIFOE3; FOXD family; 
   FIFO = TKT; CLH; MCS; 
*  LOITER = FOXD family
   MCSCR  = MCSCRA8U
   LIFOCR = LIFOE3
*  CR : mostly-LIFO admission order
*  STP = Spin-then-park waiting policy
}

\Invisible{
Candidates for the name of lock algorithm and the paper title.
Alternate names-titles and finalists :
*  Torc = Throttling with restricted-regulated circulation; 
*  Non-Nocere; 
*  Reticle = restricted-Regulated Thread Circulation; 
*  ACME = Admission control for mutexes; 
*  Crux = Concurrency Regulation-Restriction Over mutexes;
*  Cortex = Concurrency Regulation-Restriction over mutexes;
*  Curated Mutexes;
*  Malthusian Mutexes; 
*  Malthusian Locks; 
*  Malthusian Concurrency Restriction via Locks;
*  Regulus; 
*  Tourniquet locks;
*  Constrictor Locks;
*  Venturi Effect Locks; 
*  Redactive or Confinement locks;
*  Parsimony
*  Sparsimony
} 

\Invisible{
*  CR = Concurrency Restriction or regulation   
*  Admission control; regulated admission; constrained concurrency; 
*  Modulate; Moderate; 
*  We propose;
*  locks are Soviet-era technology; 
*  Shih: computation; communication; caches 
*  Venturi Effect : restrict flow implies reduce pressure and faster flow velocity; 
*  Performance diode -- only improves; never degrades; 
*  http://www.newyorker.com/magazine/2011/02/07/crush-point 
*  MPL = Multiprogramming level; 
*  Tragedy of the commons = rational maximizing behaviors by
   individuals results in unsustainable overexploitation of a resource. 
*  Performance is subadditive 
*  Slower-is-Faster phenomenon 
*  Concurrency control overheads are ultimately proportional to
   contention instead of actual throughput. 
*  Collection of institutional knowledge of interest to practitioners 
*  Amdahl's law vs Gunther's Universal Scalability Law - USL 
*  Parameter parsimony
}

\Invisible{Oracle Invention Disclosure Accession number : ORA150878} 
\Invisible{Extended version of EuroSys 2017 paper} 

\end{abstract}

%% \bigskip
%% \centerline{{\bf Keywords}: mutual exclusion; mutex; local spinning }
%% \end{titlepage}
%% \category{D.4.1}{Operating Systems}{Process Management}{Concurrency}
%% \category{D.1.3}{Concurrent Programming}{Parallel Programming} 


%=========================================================================
%  Introduction
%=========================================================================

\section{Introduction}

The scaling collapse phenomenon mentioned above arises variously from communication and
coordination overheads or from competition for any one of a number of shared resources. 
This paper focuses on the latter -- we explore the etiology of
scaling collapse via resource competition in more detail below.  For example, one such 
resource is the shared last-level cache (LLC) on a single-socket system.   
All the cores on the socket compete for residency in the LLC, and concurrent requests 
from those cores may cause destructive interference in the LLC, continuously eroding 
the residency of the data from any one core.  

\Invisible{
*  Destructive interference; fratricide; internecine; pyrrhic; 
*  Impede; impedance; impediment 
*  Residency; occupancy; tenancy;} 

\Invisible{Typically we think of resource competition as zero-sum game, but 
in practice it can be a negative sum game.} 

The effect is similar to that of \emph{thrashing} 
as described in Denning's \emph{working set} model of memory 
pressure \cite{denning-workingsets}. 
A system is said to thrash when memory is overcommitted and the
operating system spends an inordinate amount of time servicing page faults,
reducing overall progress. 
The solution in that context is \emph{swapping} -- the transient deactivation 
of some subset of the concurrently running programs. 
The medium-term scheduler responds to excessive paging and potential thrashing 
by swapping out selected ``victim'' processes until the thrashing abates.  
This closely models our approach where we transiently deactivate
excess contending threads that do not contribute to improved throughput.  
CR responds to contention instead of memory pressure.  We extend Denning's 
%% Wording: by analogy we extend ...
ideas from memory management to locks, defining the \emph{lock working set} (LWS) 
as the set of distinct threads that have acquired a given lock in some time interval.  
%% For our purposes 
We use the ordinal acquisition time of the lock to define the 
interval instead of wall-clock time.  Suppose threads $A$, $B$, $C$, $D$ and $E$
contend for lock $L$ and we have an admission order (also called the \emph{admission history}) 
of $A$ $B$ $C$ $A$ $B$ $C$ $D$ $A$ $E$ for admission times $0-8$, respectively.  
%% We call this the admission history.  
The LWS for $L$ for the period $0-5$ inclusive is threads $A$ $B$ $C$ and the 
\emph{lock working set size} (LWSS) for the period is thus 3 threads.  

\Invisible{
* Logical time; logical acquisition time; 
* Throughput; progress;
* Selected processes; victimize;
* paging devolves to swapping under sustained pressure.
* reduce; minimize; 
* admission hhistory; admission order; acquisition order; 
* http://web.stanford.edu/~ouster/cgi-bin/cs140-winter12/lecture.php?topic=thrashing
}

%% CONSIDER: move this following para to "Evaluation" section.  ...
%% This para discusses short-term fairness (average LWSS and MTTR) while para in 
%% eval section discusses long-term fairness (Gini and RSTDDEV of steps  completed).  
%% It's important to quantify the degree of fairness in order to
%%
%% better gauge the trade-off between throughput and fairness.  
%% we quantify both short- and long-term fairness.  
%% Say we have the acquisition history of a lock over some measurement period.   
%% We decide on a LWSS window size $W$ and partition the history into $W$-sized disjoint 
%% windows, compute the LWSS of each window, and then take the average of those values.  
%% Given an acquisition history for a lock over some measurement period we partition the
%% 
%% Average LWSS or simply LWSS if the window size is implied.  
%% Implied by context

CR may be unfair over the short-term, but our admission policies intentionally impose
long-term fairness \footnote{Fairness measures how admission order deviates from 
arrival order or from strict FIFO order.}. 
To help gauge the trade-off between throughput and fairness we introduce two metrics for
short-term fairness.  
For the first metric, we partition the admission history of a lock 
into $W$-sized disjoint abutting windows, compute the LWSS of each window, and take the average
of those values.
We refer to this value as the \emph{average LWSS} over the measurement interval -- 
it gives an intuitive measure of short-term fairness. 
In this paper we use a window size of $1000$ acquisitions, well above the maximum
number of participating threads.  
The second measure of short-term fairness 
is the \emph{median time to reacquire} (MTTR), computed over the entire acquisition 
history.  \emph{Time to reacquire} is determined at admission time, and is the number of 
admissions since the current thread last acquired the lock.  
Time to reacquire is analogous to \emph{reuse distance} in memory management.   

%% \cite{blog-shorttermfairness} and \cite{blog-longtermfairness} 
%% \url{https://blogs.oracle.com/dave/entry/measuring_short_term_fairness_for}
%% \url{https://blogs.oracle.com/dave/entry/measuring_long_term_fairness_for}  

\Invisible{
*  MTTR = Median-time-to-reacquire
*  TTR is always $>=$ waiting time, measured in acquisitions. 
*  Myriad ways exist to measure short- and long-term fairness. 
*  Under an ideally fair FIFO lock, admission order corresponds perfectly with arrival order. 
*  Define unfairness as : how admission order deviates from FIFO or how admission order
   deviates from arrival order. 
*  LWSS Window size $W$ should be larger than maximum number of concurrent participating threads. 
*  Admission order; arrival order; FCS-FIFE-FIFO; fairness; schedule;
   Admission schedule; 
*  Sacrifice; trade-off;
*  Quantify short-term and long-term fairness
*  metric; measure; quantify; evaluate; characterize; figure-of-merit
} 

CR acts to reduce the number of distinct threads circulating through the lock 
over short intervals and thus tends to reduce the LWSS, while still providing long-term fairness.
The CR admission policy must also be \emph{work conserving} and never under-provision the lock.
It should never be the case that the critical section remains intentionally 
unoccupied if there are waiting or arriving threads that might enter --
if such threads exist, then one will promptly be \emph{enabled} to do so.  
\Invisible{If arriving or waiting threads might enter the critical section, then
one will be promptly \emph{enabled} to do so.} 

\Invisible{
*  The admission policy for an optimal CR implementation attempts to ..
*  If arriving or waiting threads might enter the critical section, then one 
   will be promptly \emph{enabled} to enter. 
*  Finally, the admission policy must keep the lock fully subscribed but not oversubscribed.  
*  CR attempts to minimize LWSS over short intervals and average LWSS over longer intervals.
*  CR attempts to minimize LWSS over short intervals while also providing long-term fairness.
*  maximizing fairness over long intervals.  
*  keywords: maximize; maintain; provide; sustain 
*  encourage; promote; 
*  scalability fade; scalability collapse; scalability failure; 
}

As noted above, CR partitions and segregates the set of threads attempting to circulate over the lock
into the ACS (active circulating set) and the PS (passive set) \footnote{
The ACS corresponds to the \emph{balance set} in the working set model, and the PS
corresponds to the set of swapped and inactive processes.}.  
Threads in the ACS circulate normally.  
%% Variations :
%% alternatives: providing; ensuring; populating with sufficient; provisioning; 
%% ............
%% Broadly, our CR policies act to minimize the size of the ACS while still ...
%% ............
%% (ORIGINAL) 
%% We desire to minimize the size of the ACS (and thus the LWSS) while still
%% remaining work conserving.  The goal is to have sufficient threads in the ACS to saturate 
%% the lock -- ensuring the critical section enjoys maximum occupancy -- but no more.  
%% ............
%% (redundant "ensuring") 
%% We desire to minimize the size of the ACS (and thus the LWSS) while still
%% remaining work conserving, ensuring there are sufficient threads in the ACS to saturate 
%% the lock -- ensuring the critical section enjoys maximum occupancy -- but no more.  
%% ............
%% We desire to minimize the size of the ACS (and thus the LWSS) while still
%% remaining work conserving, providing sufficient threads in the ACS to saturate 
%% the lock -- ensuring the critical section enjoys maximum occupancy -- but no more.  
%% ............
%% We desire to minimize the size of the ACS (and thus the LWSS) while still
%% remaining work conserving, ensuring there are sufficient threads in the ACS to saturate 
%% the lock and ensuring the critical section enjoys maximum occupancy, but no more.  
%% ............
%% We desire to minimize the size of the ACS (and thus the LWSS) while still
%% remaining work conserving, ensuring there are sufficient threads in the ACS to saturate 
%% the lock and that the critical section enjoys maximum occupancy, but no more.  
%% ............
We desire to minimize the size of the ACS (and thus the LWSS) while still
remaining work conserving, ensuring there are sufficient threads in the ACS to saturate 
the lock -- and that the critical section enjoys maximum occupancy -- but no more.  
Surplus threads are culled from the ACS and transferred into the PS where they remain quiesced. 
Conversely a deficit in the ACS prompts threads to be transferred from the PS
back into the ACS as necessary to sustain saturation. 
%% prompts transfer of threads ...
To ensure long-term fairness our 
approach periodically shifts threads between the ACS and PS.  
Ideally, and assuming a steady-state load, 
%% threads in the ACS will have to wait only briefly to acquire the lock and 
at most one thread in the ACS will be waiting at any moment,
reducing wait times for ACS members. 
That is, at unlock-time we expect there is typically just one thread from the ACS waiting to 
take the lock.  
%% CONSIDER: para break here ..
Intuitively, threads in the ACS remain ``enabled'' and operate 
normally while threads in the PS are ``disabled'' and do not circulate over the lock.  
Threads sequestered in the PS typically busy-wait (spin) in a \emph{polite} \cite{T4WRPause} 
fashion on a thread-local flag, or block 
in the operating system, surrendering their CPU.  
(Such polite waiting reduces the resources consumed by the waiting threads, and
may allow other threads to run faster).  
Our approach constrains and regulates the degree of concurrency 
over critical sections guarded by a contended lock in order to
conserve  shared resources such as residency in shared caches.  
Specifically, we minimize over the short term the number of distinct threads 
acquiring the lock and transiting the critical section.

\Invisible{
*  Spin; busy-wait; poll; idle; delay; wait; 
*  Conversely, threads are transferred from the PS into the ACS as necessary 
   to sustain saturation. 
*  At unlock-time if the ACS is found to have a deficit then we reprovision,
   transferring threads from the ACS from PS as necessary to ensure a work-conserving 
   admission policy. 
*  To ensure full utilization; ensure saturation; sustain saturation 
*  Even when ineffective, CR does no harm, 
   adhering to the principle of \emph{primum non nocere}.
*  Benign; 
*  Surplus threads are sequestered in the PS and remain quiesced and otherwise inactive
*  Quiesced and otherwise inactive 
*  Threads in passive set are quiesced and otherwise inactive.
*  Anti-starvation policies; languishing; impatient; 
*  Saturation; Oversaturation; Undersaturation; utilization; 
*  utilization fraction; occupancy; tenancy
*  We ensure the ACS is sufficiently large to saturate the lock, but no larger.  
*  Threads in the ACS remain \emph{hot} -- retain cache residency -- and 
   those in the PS are typically \emph{cold}.  Hot threads -- those with residency -- tend 
   to remain hot, and our policies act to keep the average ``temperature'' warmer.
   Hot CS data also tends to remain resident. 
*  ACS is typically implicit; PS is typically explicit list; 
*  CR selectively culls and passivates a subset of the threads circulating over the lock. 
*  subset of threads
*  The technique is palliative, but is often effective in practice.  
*  Variously: our approach operates by; we implement CR by 
*  Our approach operates by selective culling and passivation of excess threads 
   circulating over the lock, moving such threads from the ACS to the PS. 
*  Our approach acts to minimize the size of the ACS while still remaining work conserving.  
*  We never under-provision the ACS.  
*  holding threads in the PS in abeyance. 
*  The technique is unfair over the short-term, but may increase throughput by
   decreasing the number of distinct threads that acquire the lock within a given interval.
*  Periodically, to ensure long-term fairness, we explicitly shift threads between 
   the active and passive partitions.
*  Excess threads are quarantined in the PS.  
*  We segregate the circulating set into two partitions : the ACS and the PS. 
*  Furthermore, it is relatively simple to impose long-term fairness by 
   periodically moving threads between the ACS and PS. 
}

For instance assume a simplified execution model with 10 threads contending
for a common lock.  The threads loop as follows: acquire the lock; execute the
critical section (CS); release the lock; execute their respective non-critical 
section (NCS).  Each such iteration reflects \emph{circulation} \Invisible{or \emph{flow}}
over the lock.  In our example the NCS length is 5 microseconds and the CS length 
is 1 microsecond.  For the purposes of explication we assume an ideal lock with no 
administrative overheads.  In this case we reach saturation -- Amdahl peak speedup -- 
at 6 threads.  At any given time 1 thread is in the CS and 5 execute in their respective NCS.  
Thus under ideal CR the ACS would have 6 threads and 4 of the 10 threads would 
reside in the PS, transiently made passive.  The 6 circulating threads in the ACS would 
enjoy a round-robin cyclic admission schedule.

\Invisible{
*  onset; incipient; impending; 
*  Circulation; thruput; traffic; transit; flow; flux; passage; 
*  litotes; litotic; not uncommon vs commonly; 
*  orchestration; coordination; communication; 
*  Amdahl; Gustafson; Eyerman; Packing; 
} 

\Exclude{
In our case we are trying to avoid thrashing in the shared cache.  (A more detailed 
description of scalability collapse phenomenon and the various underlying causes 
appears below, but thrashing in shared caches is one contributing factor).  
Borrowing from the memory working set idea above, we can define the ``lock 
working set'' (LWS) as the set of threads that circulated over the lock in 
some interval [FS].  Generally, our policies strive to minimize the LWS size 
over short intervals while still keeping the lock fully saturated and subscribed.  
We accomplish this by partitioning the circulating threads into the ACS and PS.
By restricting and constraining the size of the ACS, we reduce the number of 
threads circulating over the lock in some interval.  The lock subsystem deactivates 
and quiesces threads in the PS.  Our approach constrains concurrency in order to 
protect resources such as residency in shared caches.  The technique is unfair over 
the short-term, but increases throughput.  Furthermore, it is relatively simple to 
impose long-term fairness by periodically moving threads between the ACS and PS.
} 

\Exclude{
We note that the peak throughput of a system may appear at a thread count
below the lock's saturation level.  In that cases concurrency restriction 
(CR) techniques provide neither harm nor benefit.  The collapse point and the 
saturation point are unrelated.  Our concurrency restriction technique may help 
but won't hurt.  Contended locks just happen to be a convenient and opportunistic 
vehicle with which to restrict concurrency.  Peak throughput never occurs
beyond -- at higher thread counts than -- the lock's saturation point. 
}

\section{Scalability Collapse} 

The scalability collapse phenomenon involves competition for shared hardware
resources.  A classic example is residency in a shared LLC.  As more distinct
threads circulate over the lock in a given period, cache pressure and
%% Unit time; given period
miss rates increase.  Critically, as the cache is shared,  
residency of the data accessed by a given thread decays over time due to the action 
of other concurrently running threads that share the LLC.  
The application may start to thrash in the LLC and become memory-bound.  
As the LLC miss rate rises from cache pressure, contention for the 
DRAM channels increases, making LLC misses even more expensive and
compounding a deleterious effect. 
CR can serve to reduce such destructive interference in shared caches.  
By reducing the number of threads circulating over the short term, 
we reduce cache pressure and retain residency for 
longer periods, reducing the miss rate and DRAM channel congestion.  

\Invisible{
*  Concurrency restriction reduces destructive interference in shared caches.
*  By reducing the LLC miss rate we also reduce contention for DRAM channels:
   DRAM channel congestion and competition for channel bandwidth. 
*  With a sufficiently large number of threads circulating over a contented lock,
   the throughput becomes dominated solely by the CS duration.
} 

\Exclude{
Analogy between swapping and lock-based concurrency restriction.
Classic medium-term OS scheduling policies passivates or ``swaps'' threads or 
processes when memory pressure grows high.  This avoids thrashing -- contention 
and competition for memory that results in destructive interference.  
swapping is a response to the onset of thrashing.
The same principle applies to caches. 
By passivating a thread we may reduce thrashing and destructive interference
in the cache.
Using locks is a convenient and expedient way to restrict concurrency as we
can usually determine a reasonable number of threads to run.
Swapping acts to reduce memory pressure. 
Concurrency restriction and unfairness act to reduce cache pressure
} 

\begin{figure}
%% \epsfxsize=2.20in \epsfbox{misses.eps}
%% \resizebox{....}
%% \includegraphics[angle=270,origin=c,width=16cm]{plot-avl/plot.eps}
%% \includegraphics[width=11cm, trim=2cm 3cm 0 0]{./CR-Graph.eps}
%% \includegraphics[width=10cm, trim=2cm 3cm 0 1cm]{./CR-Graph.eps}
\includegraphics[rviewport={.1 0 .9 .9 },clip,width=9.4cm,angle=0,origin=c]{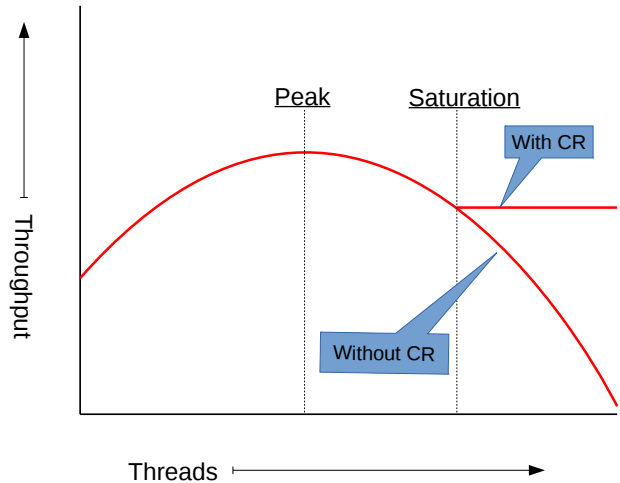} 
\vspace{-45pt}  %% tighten vertical spacing
\caption{Impact of Concurrency Restriction} 
\label{Figure:CR-Graph} 
\end{figure}
%% \vspace{-8pt} 

Figure \ref{Figure:CR-Graph} depicts the impact of CR
via an idealized aggregate throughput graph.  Thread count appears on the X-axis
and aggregate throughput on the Y-axis. 
In our depiction there are more logical CPUs than threads, so preemption is not a factor.  
Such concave scaling graphs are
common in practice, and reflect scalability collapse \cite{lsap11-cui,KunRen}  
\footnote{Lock implementations themselves are sometimes a causative factor for collapse,
for instance via induced coherence traffic on lock metadata
or where lock algorithmic overheads increase with the number of contending or participating threads.}. 
We show that a properly designed lock with CR can
also act to reduce collapse stemming from competition for shared hardware resources.  
Assume an execution model
with one contended lock $L$, where each thread repeatedly acquires $L$, executes a critical
section, releases $L$, and then executes a non-critical section.
All threads start at the same time and run concurrently throughout the measurement interval.
Throughput on the Y-axis reflects the total number of iterations completed
by the threads in the measurement interval.  
Maximum throughput appears at the threading level corresponding to \textit{Peak}, representing
the key inflection point where performance drops as thread counts increase.  Beyond
\textit{peak}, additional threads do not contribute to performance, and in fact may
degrade performance.  
This behavior is also called \emph{retrograde scaling} \cite{arxiv-0809-2541-gunther}.
\textit{Saturation} reflects the minimum threading level where there is always at
least one waiting thread when the owner releases $L$ -- the onset of sustained contention 
where the lock is expected to be held continuously
(or nearly so, for test-and-set locks in transition) 
and the critical section is continuously occupied.  
We say threads beyond
\emph{saturation} are \emph{excess} or surplus threads -- threads not necessary
to achieve saturation.  The thread count for \textit{peak} will always
be less than or equal to \textit{saturation}.  CR can begin to operate and provide benefit when 
the thread count exceeds \textit{saturation}. The value for \textit{peak} is 
imposed by platform architectural factors, overall system load, and
offered application load, and is unrelated and 
orthogonal to \textit{saturation} \footnote{Contended locks just happen to be a convenient
and opportunistic vehicle with which to restrict concurrency.}.
The value for \textit{peak} is not usually amenable to analytic calculation, and, when
required, is determined empirically.  

\Invisible{
*  arg-min; arg-max
*  unrelated; orthogonal; independent of
*  Over-threading and contention not coupled; decoupled; 
*  We use a fixed-time-report-work methodology. 
*  Saturation implies that the lock remain held and the CS remains occupied. 
*  Saturation = point of sustained contention; 
*  Retrograde scaling 
}

We note two regions of interest.  First, when the thread count is less than 
\textit{saturation}, CR would be ineffective and does not operate.  CR 
does not impact performance in this region, providing neither harm nor benefit. 
%% adhering to the principle of primum non-nocere.  
Second, when the thread count exceeds \textit{saturation}, CR can operate, 
ideally avoiding the subadditive scalability collapse evident in the graph when CR is not enabled.  
CR acts by clamping the effective thread count -- over the short term -- to 
\textit{saturation}.  Beyond \textit{saturation} and under fixed load we expect the LWSS to always be greater 
than or equal to \textit{saturation}.  
%% OPTIONAL: ...
\Invisible{ Arguably, CR does not improve performance in this region, but it avoids 
worsening performance arising from scalability collapse.}  
\Invisible{T >= saturation implies T >= LWSS >= saturation} 
\Exclude{Preserve; retain; sustain; protect; ... performance} 

%% \section{Locking algorithms that provide concurrency restriction}

\section{Taxonomy of Shared Resources} 

We provide a limited taxonomy of inter-thread 
shared  resources that are subject to competition and are amenable to conservation 
via CR.  Each of the following shared resources identifies a potential 
mode of benefit for CR.

\Invisible{
*  Etiology; underlying cause; Causation; Mode-of-benefit; 
   phenomena; phenomenon; phenomenology; effect; arise from;  
*  MoB = Mode-of-benefit
*  Identify; enumerate; list; taxonomy; 
*  Sub-additive performance ; concave; 
*  Scalability collapse phenomena and the classic concave scaling curve 
   result from competition for shared resources.
*  amenable to; subject to
*  Let thruput(n) be the thruput function for n threads
   thruput(n) vs n*thruput(1) 
}

%% CONSIDER: $\dagger$

\begin{ItemizeTight}
\item Socket-level resources
\begin{ItemizeTight}
\item LLC residency and DRAM channel bandwidth 
\item Thermal and energy headroom -- enablement of Turbo mode\cite{TurboDiaries} 
\item Intra-socket inter-core cache-coherent interconnect bandwidth
\Invisible{cache-to-cache transfers: C2C} 
\end{ItemizeTight}
\item Core-level resources
\begin{ItemizeTight}
\item Pipeline and floating point unit availability 
\item Core-private L1 and L2 residency -- cache pressure 
\item Translation lookaside buffer (TLB) residency 
%% OPTIONAL ...
\Exclude{\item Branch predictor state} 
\end{ItemizeTight}
\item System-wide resources 
\begin{ItemizeTight}
\item Logical CPUs 
\item Inter-socket NUMA cache-coherent interconnect bandwidth
%% \item Traditional system-wide memory availability 
%% \item Operating system page cache residency 
\item Memory availability and pressure -- system-managed memory residency and paging
\item I/O channel availability 
%% OPTIONAL ...
%% \item I/O channel availability
%% \item Operating system buffer-page cache residency and ``seek span''
%% CR can act to restrict size of classic working-set and paging/swapping.  
\end{ItemizeTight} 
\end{ItemizeTight}

\Invisible{
*  Coherence fabric; crossbar; invalidation diameter; number of participating
   L2 caches; intra-socket inter-core interconnect bandwidth; analogous to NUMA, but
   on-chip; channel contention and congestion; 2:1 mux that connects 
   pairs of cores to cross-bar; 
*  Other figures of merit for lock algorithms:
   Joules/op; Virtual-cpu-time/op; wall-clock-time/op; cycles/op; 
   aggregate-ops/time;  
*  Exemplar; example; illustrate; demonstrate; canonical; representative; 
   illustrative; illuminate; exhibit 
*  support; confirm; probative; 
*  intentionally selected to illustrate; contrived; demonstrate; support; probative
*  For the purposes of explication; for the purpose of brevity 
*  For each of the resources above, we can construct microbenchmarks
   that manifest destructive competition absent CR, and avoid that issue with CR. 
}

Competition for core-level resources such as pipelines typically starts to 
manifest when the number of ready threads exceeds the number of cores,
and more than one thread is running on a core.
The onset of competition for socket-level resources may start at lower thread counts.
Contention for CPUs occurs when the number of ready threads exceeds
the number of logical CPUs, where preemption (multiprogramming) starts. 

%% REDUNDANT: contended vs fully saturated
As noted previously, a key socket-level shared resource is LLC residency.
Suppose we have a contended lock that is fully saturated.  In this mode
the critical section duration solely dictates throughput \cite{isca10-Eyerman} 
\footnote{When a set of threads is \emph{contention-limited} by a common lock,
the duration of the critical section solely determines throughput, which
is insensitive to the duration of the NCS.  Assuming saturation is maintained, reducing the NCS
duration simply causes circulating threads to arrive more quickly at the lock
and to wait longer, with no improvement in throughput.  We note, however, that
more efficient NCS execution that consumes reduced resources may provide benefits 
in the case of multi-tenancy where unrelated threads -- thread not circulating 
over a commmon lock -- compete for shared resources.}.  
\Exclude{The data accessed in the non-critical sections is thread-private and
multiple threads may execute their respective NCS concurrently with the single
thread in the CS.  Multiple independent NCS instances are accessed
at the same time time as the CS data and accesses to non-critical data -- 
performed by threads executing their respective non-critical sections -- 
will displace and evict critical data}
Data accessed in non-critical sections is thread-private and multiple
independent non-critical sections may execute concurrently with a single CS.   
NCS accesses displace and evict critical data  
\footnote{CS invocations 
under the same lock typically exhibit \emph{reference similarity}: acquiring lock $L$ is a good 
predictor that the critical section protected by $L$ will access data that was 
accessed by recent prior critical sections protected by $L$.  That is,
CS invocations tend to access data accessed by prior CS invocations, exhibiting
inter-CS inter-thread locality and reuse.}.  
%% yielding classic destructive interference in the LLC.
%% destructive interference in the shared LLC.
%% Subsequent critical sections will incur more misses, and scaling will suffer.
As the set of threads circulating over the lock grows, the total non-critical footprint
increases, and we find more cache pressure in the communal LLC.  
In turn, the critical section suffers more LLC misses, \Exclude{and DRAM channel contention}
increasing the duration of the CS and decreasing throughput over the contended lock.  
CR can afford benefit in this circumstance by restricting the set of circulating
threads, reducing cache pressure and thus increasing throughput compared to a 
perfectly fair FIFO lock
%% OPTIONAL ...
\footnote{Various forms of competition for LLC residency are possible:
CS-vs-CS, NCS-vs-CS, and NCS-vs-NCS.  We assume sufficient contention that
aggregate throughput is solely controlled by CS duration, in which case inter-NCS
``fratricide'' is the least important mode.  Even so, NCS-vs-NCS competition
can increase DRAM channel contention.}.  

\Invisible{Increasing the CS duration decreases aggregate throughput over the contended lock.} 
\Invisible{We note that CS-vs-CS, NCS-vs-CS and NCS-vs-NCS competition for
LLC residency are all possible.}    
\Invisible{We note that the non-critical sections will erode and decay 
each other's residency in the LLC as well as that of the critical section data.  
We assume sufficient contention that aggregate throughput is controlled solely by 
the critical section duration.  Even so, inter-NCS ``fratricide'', 
%% while it may not directly impact throughput, 
can increase DRAM channel contention.} 
\Exclude{The non-critical data accesses erode and decay the residency
of the critical data -- we have classic destructive interference. } 
\Exclude{strict perfectly fair FIFO lock} 
%% Suppose we have; say we have; 

%% Detailed example; worked example ...
We next provide a detailed example to motivate the benefit of CR on a single-socket
SPARC T5 processor where the shared LLC (L3 cache) is 8MB. 
We have a customer database that is 1MB, and each CS operation will access a record 
in that database.  Each record resides on a single cache line. An individual CS 
will access only one record, but over time 
most records will be accessed repeatedly by subsequent operations. (The CS may be 
``short'' in average duration but ``wide'' in the sense that a sequence of CS operations 
will eventually access a large fraction of the records).  We have 16 threads, and on an otherwise unloaded
system the NCS duration is 4 times that of the CS duration. The $(NCS+CS)/CS$ ratio is such that only 5 
threads are needed to fully saturate the 
lock and provision the ACS.  Furthermore, the NCS footprint of each thread is 
1MB.   Even though an individual NCS operation might be short, over time a 
thread will access all 1MB of its thread-private data.  Recall that the CS data 
is shared and the NCS data is per-thread and thread-private.   Under a classic 
%% Per-thread; thread-private
FIFO MCS lock \cite{MCS}, all 16 threads will circulate over the lock in round-robin 
cyclic order.  The total footprint is 17MB : $(16 \ threads * 1MB/thread) + 1MB $
for the CS, exceeding the 8MB capacity of the LLC.   The 
NCS operations will erode and decay the residency of the CS data, slowing 
execution of the CS, and degrading overall throughput.  But 
with CR the lock subsystem is able to  limit the size of 
the ACS to 5 threads.   In this mode, the total short-term footprint is 6MB : 
$(5 \ threads * 1MB/thread) + 1MB$ for the CS.  The total footprint -- 
the CS data plus the NCS data of the ACS threads -- fits comfortably within 
the LLC.   Consequently, the NCS instances do not erode CS residency, the 
CS does not suffer from misses arising from destructive interference in the LLC, 
and throughput is improved.   CR reduces cache pressure and in
particularly on CS data.  ``Hot'' threads -- those that have run recently and 
have residual LLC residency -- tend to remain ``hot''.  

\Invisible{
*  By minimizing the ACS, CR can increase throughput. 
*  CS data references under a given lock will exhibit inter-CS 
   inter-thread temporal locality and reuse
}
  
%% OPTIONAL ...
%% redundant with paragraphs above 
\Invisible{
The actions by concurrent threads in the NCS will erode the LLC cache 
residency of the CS data.  If the ACS is large, the eviction pressure on 
the CS data by the multiple NCS instances becomes more intense.  In turn, the 
CS incurs more misses in the LLC; the CS duration increases, and 
throughput consequently drops.  Thus by minimizing the ACS, we can increase 
aggregate throughput. } 

\Invisible{
*  replace; Displace; erode; Decay; expel; evict; pollute; attrition; attrit;  
*  Fratricide; internecine; 
*  Congestion; Contention; Competition; Conflicts 
*  Little's law; PASTA property} 

\Invisible{Amdahl's law is not a perfect match for locking.  Amdahl's model assumes
a serial phase where no threads run -- or just 1 thread runs -- alternating with
a pure parallel phase where all thread can this.  Arguably, Amdahl models barriers, 
although barriers still allow some concurrency when some threads have arrived 
for rendezvous and other threads are still running.  Amdahl does not faithfully
model locking.  When a thread holds the lock, other threads can concurrently 
execute their NCS regions. See Eyerman.}  

%% TAG:SLEEP
Another socket-level shared and rationed resource is thermal and energy headroom.  
By running fewer threads in a given interval relative to other locks, CR may 
reduce energy use and heat dissipation.  
Furthermore, by quiescing threads in the PS and allowing more processors to enter and
remain in deeper low-power \emph{sleep states} while idle, our approach can enable 
\emph{turbo mode} \cite{TurboDiaries, TurboEnablementPatent} for the remaining 
active threads -- critically including the lock holder -- accelerating their 
progress and improving throughput. 

\Invisible{In Praise of Idleness, by Bertrand Russell} 

%% OPTIONAL ...
\Invisible{Space considerations do not allow us to present the details, but we
used the \textbf{RAPL} -- Running Average Power Limit -- facility on modern
Intel processors to measure Joules consumed and the enablement of turbo mode.
We also used the RAPL facility to constrain power usage. 
In power-constrained configurations, CR provided even higher relative benefits.}  
\Invisible{Spinning expends energy, which is rationed.} 
\Invisible{This paper focuses on socket-level LLC residency, core-level cache 
and DTLB residency, core-level pipeline availability, and system-wide logical 
CPU availability.} 

The \emph{waiting policy} of a lock implementation (discussed below) defines how a 
thread waits for admission, and can have a significant impact on 
competition for core-level resources such as pipelines, socket-level resources such 
as thermal and energy headroom, and global resources such as logical CPUs.  

\Invisible{Keywords: impact; influence; interaction; interplay} 

\section{The MCSCR lock algorithm} 

We now describe the implementation of \textbf{MCSCR} -- a classic MCS lock \cite{MCS} modified
to provide CR by adding an explicit list for members of the PS 
\footnote{Under classic MCS, arriving threads append an element to the tail of the
list of waiting threads and then busy-wait on a flag within that element.
The lock's \texttt{tail} variable is explicit and the head -- the current owner -- is implicit. When the
owner releases the lock it reclaims the element it originally enqueued and sets the flag
in the next element, passing ownership.}.   
At unlock-time, if there exist any intermediate
nodes in the queue between the owner's node and the current tail, then we have
surplus threads in the ACS and we can unlink and excise one of those nodes and 
transfer it to the head of the passive list where excess ``cold'' threads reside.
This constitutes the culling operation. 
\Invisible{
Specifically, our approach simply looks forward into the MCS chain to detect
the onset of contention.} 
Conversely, at unlock-time if the main queue is empty 
except for the owner's node, we then extract a node from the head of the passive list, insert 
it into the main queue at the tail, and pass ownership to that thread, effectively transferring
an element from the PS back into the ACS.  
This ensures MCSCR is work conserving and provides progress and liveness. 
The element at the head of passive list is the most recently arrived member of the PS.  
Absent sufficient contention, MCSCR operates precisely like classic MCS.  
MCSCR directly edits the MCS chain to shift threads back and forth 
between the main chain and the explicit list of passivated threads
\footnote{Editing the MCS chain was first suggested by Markatos et al. \cite{markatos} for
the purposes of enforcing thread priorities.}. 
The ACS list is implicit, while the PS -- the excess list -- is explicit. 
MCSCR detects contention and 
excess threads simply by inspecting the main MCS chain.  

\Invisible{
*  If there are any intervening threads between the tail and the owner's node, then 
   those threads are excess and can be culled.
*  3 or more nodes, including owner's node, on the MCS chain
*  Intervening; intermediate; excess; surplus; 
*  We now describe an implementation of CR based on the classic MCS lock.  
   We modified the classic MCS lock \cite{MCS} by adding an explicit list of 
   passivated excess threads, yielding the \textbf{MCSCR} lock.   
*  We note that CR can be easily applied to wide variety of existing locks. 
}
 
To ensure long-term fairness, the unlock operator periodically selects the tail $T$
of the PS as the successor and then grafts $T$ into the main MCS chain
immediately after the lock-holder's element, passing ownership of the lock to $T$.
Statistically, we cede ownership to the tail of the PS -- which is the least
recently arrived thread -- on average once every 1000 unlock operations.  
We use a thread-local Marsagalia xor-shift pseudo-random number generator 
\cite{Marsaglia-xorshift} to implement Bernoulli trials
%% to control passing ownership to the tail of the excess list.  
The probability parameter is tunable and reflects the trade-off between  
fairness and throughput.  
Transferring a thread from the PS into the ACS typically results in some other member 
of the ACS being displaced and shifted into the PS in subsequent culling operations. 

\Invisible{
*  Degenerate case where we always pull from PS is just normal MCS.
*  Homeostatis; converge to steady-state
*  Use Bernoulli trials to trigger tail extraction.
*  Desirable state; preferred; target
} 

Culling acts to minimize the size of the ACS. Under fixed load, aggressive 
culling causes the system to devolve to a desirable state where there is at most one 
member of the ACS waiting to acquire the lock.  In this state, the ACS 
consists of that one waiting thread, the current owner of the lock, and a 
number of threads circulating through their respective non-critical sections.
The size of the ACS is determined automatically and is not a tunable parameter. 
At unlock-time, the owner will usually pass ownership of the lock to that waiting thread.
\Invisible{The waiting thread will typically take the lock after the owner releases it.} 
Subsequently, some member of the ACS will complete its non-critical section
and wait for the lock.  In this mode, admission order is effectively cyclic 
over the members of the ACS. \Invisible{and mostly-LIFO in general}

\Invisible{
*  regardless of the prevailing lock admission policies.  
*  The MCS lock protects the excess list.
} 

%% OPTIONAL : 
All changes to support MCSCR are implemented in the unlock path; the
MCS lock operator remains unchanged.
Operations on the PS occur within the unlock operator while the MCS lock
is held -- the PS is protected by the MCS lock itself. This artificially increases the 
length of the critical section, but the additional manipulations are short and
constant-time. 

\section{Lock Design Fundamentals} 

\Invisible{We describe a number of lock implementation properties that influence our
design decisions.} 

%% Microeconomics; synchronization fundamentals
%% Fundamentals that influence lock design
%% Lock Design Fundamentals; Principles; Precepts; 
%% criteria and variations
%% Possibly decompose-deconstruct : 
%% * HW-level facilities and factors
%% * OS-level facilities and factors

\Invisible{
*  Precepts; Principles; Fundamental; Variations 
*  Background; explanatory; supplemental; supplementary; 
*  Lock Microeconomics; 
*  criteria; desiderata; 
*  Economics; cost model; profitable; 
*  costs that inform and influence design;} 

\subsection{Waiting Policies} 

\Exclude{
A classic MCS or ticket lock provides succession by direct handoff.  The owner 
directly transfers the lock to some pending successor.   A barging lock  
competitive succession drops the lock, and, if necessary, enables 
some pending threads to recompete for the lock.   Almost everything in the 
JVM uses barging.   The linux and solaris pthread\_mutex is barging by default.    
With regards to waiting policy (spin, block, spin-then-block), direct succession 
and FIFO in particular are disastrous with block or spin-then-block.   On an 
unloaded SPARC Solaris T4-1, the latency from the start of an unpark to the 
time the target resumes is about 14K cycles, best case.   The unpark itself 
takes about 3K cycles to return.   
And we can't allow unbounded spinning, so direct succession tends to leave us in a 
painful position.   Barging also tends to reduce convoying.   In particular, barging avoids the 
case where we pass ownership to a preempted thread.  } 

The choice of waiting policy used by a lock implementation influences competition
for CPUs, pipelines and thermal headroom, making the selection of a waiting policy
critical for CR.  The waiting policy also dictates key latencies, further informing
our design. We identify a number of commonly used policies:

\Invisible{ The manner in which a thread waits for a lock is the waiting policy.}

%% Variations for list of waiting policies :
%% @  Explicit list
%%    \begin{ItemizeTight}
%%    \item\textbf{Unbounded spinning} \\
%% @  \subsection{Unbounded Spinning}
%% @  \vspace{2mm} \noindent \textbf{Unbounded Spinning} \vspace{1mm} \\ \noindent
%%    Perhaps also use \hfill

\vspace{0.5 \baselineskip} 
%% \vspace{2mm} 
\noindent
\textbf{Unbounded spinning} \vspace{1mm} \\
\noindent 
Classic MCS and test-and-set spin locks (TAS locks)\cite{anderson-spinning} use unbounded spinning, also
called busy-waiting or polling.  Waiting threads simply loop, re-checking
the variable of interest.  
While unbounded spinning appears often in academic literature, actual deployed
software generally avoids indefinite spinning.  At some point a spinning
thread is expected to deschedule itself.   While convenient and simple,
unbounded  spinning can interfere with the performance of other threads on 
the system by consuming pipeline resources.  Spinning also expends energy and consumes available
thermal headroom, possibly to the detriment of sibling cores that might otherwise enjoy
turbo mode acceleration.  In addition, a spinning thread occupies a processor, possibly prohibiting 
some other ready thread from running in a timely fashion. 
(In fact spinning threads might wait for the lock holder which has itself been preempted.) 
If there are more ready threads than logical CPUs, then preemption by 
the kernel would eventually ensure those other threads run, but those ready threads may 
languish on dispatch queues until the spinners exhaust their time slice. 
Typical quanta durations far exceed the latency of a voluntary context switch.  
Despite those concerns, spinning remains appealing because it is simple and
the lock handover latency (discussed below) -- absent preemption -- is low.  
%% appealing vs popular

\Invisible{If a thread uses unbounded spinning then 
eventually involuntary preemption by the operating system will deschedule the 
spinner and allow those other ready threads to run, but quanta (time slice) can be 
relatively long, so depending on preemption is not prudent and can result in 
particularly poor performance when the number of ready thread exceeds the number 
of available processors.} 

\Invisible{
*  Spinning; Busy-waiting; polling; active waiting
*  Blocking; passive waiting; deschedule; suspend; sleep; 
*  Visible vs invisible spinners; publish existence of waiting thread
*  Pause is like hardware yield primitive.
} 

Spinning can be made more \emph{polite} to sibling threads by using the \texttt{PAUSE} 
instruction on x86, or the \texttt{RD CCR,G0} idiom, a long-latency no-op, on SPARC.
These instructions transiently cede pipeline resources to siblings -- logical CPUs that share 
the core with the spinning thread -- allowing those siblings to run faster
\footnote{When only one logical CPU is active in a core, the per-core pipelines automatically 
fuse and provide better performance for the single active CPU.  
Intel processors with \emph{hyperthreading} exhibit similar behavior.  
Polite spinning via the \texttt{WRPAUSE} instruction or the \texttt{RD CCR,G0} 
idiom also enables fusion.}. 
Such instructions may also reduce power usage.

%% OPTIONAL ...
SPARC also provides the \texttt{WRPAUSE} instruction with a parameterized delay period
\cite{T4WRPause}. 
%% OPTIONAL ...
Longer pauses yield more benefit to siblings but may impact response latency by 
creating ``dead time'' and lag when ownership is passed to a waiting thread that 
happens to be in the middle of a WRPAUSE operation.  This presents an altruism 
trade-off: longer delays are more polite and provide more benefit to siblings, 
but may also increase lock handover latency. 

\Invisible{
*  Polite spinning and parking reflect altruism  
*  WRPAUSE useful for short-term waiting 
*  WRPAUSE is useful for both global and local spinning
} 

\Exclude{
The WRPAUSE instruction on modern SPARC processors cedes the pipelines to sibling 
strands.   This is politeness with respect to local resources.    By using WRPAUSE we also may (a) 
use less energy if the system is uncapped, or, (b) if the system is thermal- or 
energy-limited or capped possibly improve overall performance by allowing the 
other threads to run faster by leaving more headroom under the cap.    For WRPAUSE 
we could use either back-off or just a fixed duration pause.    Longer pauses are 
more polite to siblings -- the spinners waste less pipeline and energy -- but we 
incur some lag when ownership passes to a thread that’s stuck in WRPAUSE.  
}

%% EXCLUDE; OPTIONAL; or include as footnote ...
The MWAIT instruction, available on x86 \footnote{Intel's MWAIT instruction is not currently available in
user-mode, impacting its adoption.} and SPARC M7 systems, allow a thread to wait 
politely for a location to change.  MWAIT ``returns'' promptly after a modification of
a monitored location.  While waiting, the thread still occupies a CPU, but 
\texttt{MWAIT}\cite{cluster13-akkan,blog-mwait} may allow the CPU to reach deeper sleep states.  
It also frees up pipeline resources more effectively than WRPAUSE.  Latency to enter and 
exit MWAIT state is low, avoiding the trade-off inherent in picking WRPAUSE durations.
Transferring ownership for locks that use local spinning is efficient and incurs little handover latency. 
MWAIT also avoids branch mispredict stalls that are otherwise inherent in exiting wait loops.  
MWAIT with a parameterized maximum time bound allows hybrid forms where a thread 
initially uses MWAIT but then falls back to parking.  MWAIT is tantamount to spinning, 
but more polite and preferred when the instruction is available.  

%% OPTIONAL 
\Invisible{MWAIT is well-suited for local spinning and we prefer MWAIT over WRPAUSE
where available.  
Using MWAIT for global TAS spinning is less clear.  
Absent additional randomization, all waiting threads will resume from MWAIT and try
the lock, likely generating futile coherence traffic.
Normally MWAIT is inappropriate for global spinning with a large number
of threads, but our approach constrains the number of threads spinning
on a given lock at any moment, making MWAIT a viable choice.  
We can also use hardware transactional memory to wait politely via TXPAUSE.}  

%% OPTIONAL ...
%% polling with sleep/yield is augmented busy-waiting
%% Threads can also attempt to voluntarily surrender their CPU -- in a polite fashion,
%% avoiding dependencies on longer-term involuntary preemption -- by calling
%% \texttt{sched\_yield} or \texttt{Sleep($D$)} where $D$ is a duration to sleep.

A busy wait loop can also be augmented to voluntarily surrender 
the waiting thread's CPU in a polite fashion 
-- avoiding dependence on longer-term involuntary preemption -- 
by calling \texttt{sched\_yield} or \texttt{Sleep($D$)} where $D$ 
is a duration to sleep.  \texttt{Sched\_yield} attempts to transfer the CPU to 
some other ready thread while keeping the caller ready.
\texttt{Sleep($D$)} makes the caller ineligible to run for the duration specified
by $D$, making the caller's CPU available to other potentially ready threads.  
\texttt{Sleep} serves to reduce the number of ready threads whereas yield does not.
Polling via \texttt{sleep} and \texttt{sched\_yield} avoids the
need to maintain explicit lists of waiting threads, as is required by the park-unpark facility
(described below).
Both \texttt{sleep} and \texttt{sched\_yield} can be wasteful, however, 
because of futile context switching where a thread resumes to find the lock remains held.  
Furthermore the semantics of \texttt{sched\_yield} are extremely
weak on modern operating systems: yield is advisory.  Spin loops augmented with
yield degenerate to an expensive form of busy waiting which is unfriendly siblings.  
The choice of $D$ presents another trade-off between
response time and politeness.  Finally, $D$ values are often quantized on modern 
operating systems, providing only coarse-grained effective sleep times. 
In practice, we find yield and sleep perform worse than simple parking.

%% OPTIONAL ...
\Invisible{Yield and sleep can put undue stress on the kernel scheduler and the 
kernel timer facility, as well precluding access to deeper energy-saving hardware
sleep states.} 

%% OPTIONAL
\Invisible{Spinning can also be augmented with \texttt{sched\_yield} calls, but these are advisory
and in practice we find they perform worse than parking.} 

%% OPTIONAL ...
\Invisible{$D$ values are quantified to the units of clock tick
interrupt periods on many operating systems -- typically between 1 and 10 milliseconds -- meaning
that short periods can not be expressed.}

%% OPTIONAL ...
\Invisible{
Spinning can be augmented with operating system \texttt{sched\_yield} calls that attempt
to yield the CPU to other ready threads.  Unfortunately sched\_yield is advisory
and has almost no semantics.  Often, it does nothing even when other runnable threads
are available on other dispatch queues.  Empirical results with \texttt{sched\_yield} are
dismal.  We do not consider it further.  
}

%% OPTIONAL
Spinning policies are further determined by the choice of local spinning
versus global spinning.  A simple fixed back-off usually suffices for local 
spinning, while randomized back-off is more suitable for global spinning. 

%% We presume the availability ...
%% We depend on the availability of a 
%% Words: use; employ; utilize; assume; incorporate; depend on; 

\vspace{0.5 \baselineskip} 
\noindent
\textbf{Parking} \vspace{1mm} \\
\noindent
%% OPTIONAL ... 
%% TAG:SLEEP
Our lock implementations employ a \emph{park-unpark} infrastructure for 
voluntary context switching.  The park-unpark facilities allows a waiting
thread to surrender its CPU directly to the operating system while the
thread waits for a contended lock.  The \emph{park} primitive blocks the caller, 
rendering itself ineligible to be scheduled or dispatched onto a CPU.
\Invisible{The thread ``deschedules'' itself via park.} 
A corresponding \emph{unpark($T$)} system 
call wakes or resumes the target thread $T$, making it again ready for dispatch 
and causing control to return from park if $T$ was blocked.
An unpark($T$) operation can occur before the corresponding park call by $T$, in 
which case park returns immediately and consumes the pending unpark action.   
Waiting for a lock via parking is polite in the sense that the waiting thread
can make its CPU immediately available to other ready (runnable) threads. 
\footnote{Threads can also wait via unbounded spinning -- busy-waiting.  In this case involuntary
preemption by the operating system will eventually make sure other ready thread will run.  
However time slices can be long, so it may take considerable time for a ready thread 
to be dispatched if there are no idle CPUs.  Parking surrenders the caller's CPU 
in a prompt fashion.}.
%% OPTIONAL ...
The Solaris operating systems exposes \emph{lwp\_park} and \emph{lwp\_unpark} system calls while 
the \emph{futex} facility can be used to park and unpark threads on Linux.  
The park-unpark facility is often implemented via a restricted-range semaphore, 
allowing values only of 0 (neutral) and 1 (unpark pending).  
The park-unpark interface moves the decision of \textit{which} thread to wake
out of the kernel and into the user-space lock subsystem, where explicit lists of
parked threads are typically maintained.

Parking suspends the calling thread and voluntarily surrenders 
the CPU on which the caller ran, making it immediately available to run other
ready threads.  If no other threads are ready, then the CPU may become
idle and be able to drop to deeper sleep states, reducing power
consumption and potentially enabling other ready threads on the same chip to run
at faster speeds via turbo mode  
\footnote{Turbo mode is controlled directly by hardware instead of software and requires sufficient
energy headroom to be enabled.  Software indirectly influences the availability of turbo
mode via waiting policies.}.
Parking also reduces competition for intra-core pipeline resources, and promotes fusion.
In turn, other threads -- possibly including the lock holder 
running in its critical section -- may run faster, improving scalability.  
%% WORDING: allows; gives opportunity; 
Parking also allows the operating system to rebalance the set of running
threads over the available cores via intra-socket migration.  This is 
particularly useful for CR \footnote{If the operating system did \emph{not} rebalance
then we could easily extend CR to itself balance the ACS over the cores, intentionally
picking ACS members based on where they run.}. 
Spinning does not allow such redistribution.
Parking also reduces the number of concurrently ready threads, in turn reducing
involuntary preemption by the operating system.  
However the costs to enter and exit the parked state are high and
require operating system services.  Thus our policies strive to reduce the
rate of voluntary context switching.  

\Invisible{
*  Park; deactivate; self-suspend; idle; block; deschedule; sleep; waiting; 
*  Parking constitutes voluntary context switching.
*  The operating system can not differentiate spinning threads vs.
   those actively working.
*  Parking provides more thermal and energy headroom for other running threads. 
*  Deeper sleep states confer more turbo benefit but also take longer to enter
   and exit, possibly increasing the latency of unpark operations.
*  Exiting those deeper sleep states incurs more latency for threads being unparked. 
*  CPUs that idle longer may reach deeper sleep states.  
}

%% TAG:SLEEP
%% OPTIONAL - REDUNDANT
\Invisible{
Parking vacates the processor, allowing other ready threads to run on that CPU and
making the caller ineligible for dispatch.  
If parking causes the CPU to become idle, the CPU can enter deeper sleep states, which in
turn confers benefit to other running threads via turbo mode. }

\Invisible{
Modern CPUs support special hardware sleep states for idle CPUs.  Deeper sleep
states draw less power and dissipate less heat, and allow more aggressive turbo 
mode for sibling CPUs on the same socket, permitting  threads on those CPUs to enjoy 
faster execution.  
Deeper sleep states may also and may promote more aggressive turbo mode for sibling cores.
Turbo mode is not controlled directly by software. Rather, it is automatically controlled 
by the CPU itself and requires sufficient energy headroom to be enabled.}  

CPUs transition to deeper (lower power) sleep states the longer they remain idle.  
Deeper sleep states, however, take longer to \Exclude{both} enter and exit. Exit latency
significantly impacts unpark latency -- the time between an unpark($T$) operation 
and the time when $T$ returns
from park.  Deeper sleep states, while useful for energy consumption and
turbo mode, may also increase the time it takes to wake a thread.  
To effectively leverage the benefits of deeper sleep states, the CPU needs to 
stay in that state for some period to amortize the entry and exit costs.  
Frequent transitions between idle and running states also attenuates the turbo
mode benefit for sibling CPUs as the CPU may not idle long enough to reach deeper states.
Lock implementations that act to reduce thread park-unpark rates will also reduce 
CPU idle-running transitions and will incur less unpark latency -- by avoiding 
sleep state exit latencies --  and also allow better use of turbo mode. 
By keeping the ACS stable and minimal, CR reduces the park-unpark voluntary
context switch rate, and in turn the idle-running CPU transition rate. 

\Invisible{
Frequent transitions between idle and running  incurs extra unpark latency
and attenuates the turbo mode benefit for siblings CPUs.  
Reducing thread park-unpark rates also reduces CPU idle-running transitions and 
concerns associated with deeper sleep states. 
As such, we our locking policies prefer to avoid frequent transitions between 
idle and running states for CPUs. }

\Invisible{
*  Lag in dispatching threads when the CPU exits idle state. 
*  Generally, the deeper the sleep state, the more power conserved while in 
   that state, but the longer it takes the CPU to enter and exit that state. 
*  Reduced context switching rates can lead to reduced CPU transitions between 
   idle and non-idle states, allowing deeper sleep states and less transition 
   overheads.} 

\Invisible{
Park-unpark and waiting via local spinning typically requires 
the lock algorithm to maintain explicit lists of waiting threads.  Generally,
it is easy to convert a lock that uses local spinning to use park-unpark or
spin-then-park.  Parking is point-to-point by nature and requires a visible 
list of waiting threads.  

Park() admits spurious returns.  A good litmus test of proper and safe 
park-unpark usage is to consider the degenerate but legal implementation 
where park() and unpark() were implemented as no-ops, in which case the 
algorithms that use park-unpark would simply degenerate to spinning.  This 
reflects a legal but poor quality implementation.  After returning from
a park() call, the caller is expected to re-evaluate the conditions related
to waiting.  Park-unpark can be thought of as an optimized from of busy-waiting
or polling.  Specifically, control returning from Park() does not imply
a corresponding previous unpark() operation.  By allowing spurious wakeups
we afford more latitude to the park-unpark implementation, possibly enabling
useful performance optimizations.  

Optimized park-unpark implementations can often avoid calling into the kernel.
Say thread $S$ calls unpark($T$) where $T$ is not currently parked.  The unpark($T$) 
operation will record the available ``permit'' in $T$'s thread structure and
return immediately without calling into the kernel.  When $T$ eventually calls 
park, it will clear that permit flag and return immediately, again without
calling into the kernel.  Redundant unpark($T$) operations -- where a waiting 
thread $T$ has previously been unparked but has not yet resumed -- also
have an optimized fast path to avoid calling into the kernel.  The only case
that requires calling the kernel is where an unpark follows the corresponding
park operation.  

Optimized Park() implementations may spin briefly before reverting to blocking
in the kernel.  The spin period is brief and bounded, and acts to reduce the 
rate of expensive and potentially unscalable calls into the kernel to perform 
ready-blocked state transitions.  This is the so-called spin-then-block waiting 
policy.  The spin period reflects local spinning and can be implemented with a 
``polite'' busy-wait loop or via MONITOR-MWAIT instructions.  

Waiting in the kernel via blocking or via MONITOR-MWAIT on a thread-private 
local variable can free up pipeline resources or bring the CPU under 
thermal-energy caps, which in turn can accelerate the progress of the 
lock owner, increasing scalability.  Recall that if the lock is contended
and fully saturated, throughput is completely determine by the critical section
duration.  By potentially accelerating the lock owner, we may reduce the
critical section duration and lock hold time. 

As noted above Park-Unpark incurs latencies for both the thread parking 
and the thread calling unpark.  If the corresponding unpark occurs after
the park operation, and the parked thread was blocked, there is 
considerable latency required to transition the wakee from blocked to 
ready to running.  

park-unpark : Appropriate for longer-term waiting; 
relinquishes to other potentially ready threads
Gives the kernel an opportunity to balance active ACS threads
over the cores and pipelines via intra-node migration.
Provides relief for sibling threads that are running on the same cores.
Reduces competition for pipelines.

To help reduce handover latency, we can use ``anticipatory warmup'' as 
follows. If we expect to unpark() thread T in the near future and T
is blocked in the kernel, then we can preemptively unpark(T) so T
becomes ready and starts spinning.  An Unpark(T) operation can impose
considerable latency in the caller because of the need to invoke kernel 
operations.  As such, an anticipatory unpark(T) should be executed while
the caller does not hold the lock for which T waits, otherwise we risk
artificially increasing the critical section length and impacting 
throughput over the contented lock.   Anticipatory unpark() operations
are particularly well suited for locks that use succession by direct
handoff, and acts to increase the odds that an unlock() operation will
transfer control to a thread that is spinning, instead of a thread that
is blocked in the kernel.  This optimization is optional, but helps
to reduce lock handover latency.
}

\vspace{2mm} 
\noindent
\textbf{Spin-Then-Park} \vspace{1mm} \\
\noindent
To reduce the impact of park-unpark overheads, lock designers may opt to use a hybrid two-phase
spin-then-park strategy.  Threads spin for a brief period -- optimistically waiting -- 
anticipating a corresponding unpark operation and then, if no unpark has occurred, 
they revert to parking as necessary.  
The maximum spin period is commonly set to the length of a context-switch round trip.  
A thread spins for either the spin period or until a corresponding unpark is observed
%% OPTIONAL ...
\footnote{Spinning can be further refined by techniques such as \emph{inverted schedctl}
\cite{blog-invertedschedctl} which reduces the impact of preemption on spinning.
The spinning period can also be made adaptive, based on success/failure
ratio of recent spin attempts \cite{dice2009adaptive}.}
%% OPTIONAL ...
\footnote{As a thought experiment, if parking and unparking had no or low latencies, then
we would never use spinning or spin-then-park waiting strategies, but would
instead simply park in a prompt fashion.  Spinning is an optimistic attempt or bet
to avoid park-unpark overheads.  Parking and spinning both reflect wasted 
administrative work -- coordination overheads -- that do not contribute directly 
to the forward progress of the application.  Spinning is arguable greedy, optimistic
and opportunistic, whiling parking reflect altruism.}.    
If no unpark occurs within the period, the thread deschedules itself by 
blocking in the kernel. (Unparking a thread that is spinning or otherwise not blocked
in the kernel is inexpensive and does not require calling into the kernel).  
Karlin et al. note that spinning for the length of a context switch and
then, if necessary, parking, is 2-competitive \cite{sosp91-karlin,tocs93-lim}. 
%% relative to ideal or optimal waiting
The spinning phase constitutes local spinning.
If available, the spin phase in spin-then-park can be implemented via MWAIT 
\footnote{Spin-then-park waiting further admits the possibility
of \emph{anticipatory warmup} optimizations where the lock implementation unparks 
a thread in advance, shifting it from parked state to spinning state.  
The lock might also favor succession to spinning threads over parked threads.}.  
%% OPTIONAL ...
We prefer parking -- passive waiting -- over spinning -- active waiting --
when the latencies to unpark a thread exceed the expected waiting period.  
\Invisible{More precisely, a thread spins until $I$ steps have passed or until
an unpark occurs.  $I$ can be expressed in either units of wall-clock time
or iterations of a spin loop.} 

\Invisible{
*  As such, we try to minimize these administrative costs.
*  Gedankenexperiment 
*  Amortize; optimistic; opportunistic; speculative; bet; gamble;
   Anticipate;
*  Losing proposition; profitable; productive; desist; 
}

%% CONSIDER: organize paper as STP and then embellishments ...
%% Possibly move STP embellishment list into appendix. 

\Invisible{
Having said that, even simple spinning — as a part of a spin-then-park waiting 
strategy — with a duration of the round trip context switch time is 2-competitive.   
This policy is pretty reasonable in practice as well as theory.  There are lots 
of minor embellishments to spin-then-park : schedctl to avoid lock waiter 
preemption; inverted schedctl; schedctl to avoid lock holder preemption; schedctl 
to avoid waiting on an OFFPROC owner;  you can make the spin duration adaptive
based on recent spin success/failure to further reduce futile spinning; 
clamp the number of concurrent spinners; bail out of the spin phase if there’s 
sufficient traffic or failed atomics; MWAIT; Polite waiting with PAUSE and WRPAUSE; 
directed yield, etc.

You can also impose concurrency restriction at a higher level outside the
waiting mechanism. 
} 

%% Possibly move this para into waiting policies | spin-then-park section
Hybrid spin-then-park \cite{dice2009adaptive} waiting strategies may reduce the rate 
of voluntary blocking and
provide some relief from such voluntary context switching costs.  However 
spin-then-park tends not to work well with strict FIFO queue-based locks.  
The next thread to be granted the lock is also the one that has waited the longest, 
and is thus most likely to have exceeded its spin duration and reverted to parking  
in which case the owner will need to be unparked, significantly lengthening the critical section
with context switching latencies.  
Spin-then-park waiting favors a predominantly LIFO admission policy.  
Generally, a waiting strategy that parks and unparks
threads is inimical to locks that use direct handoff, and to FIFO locks specifically.  

\Exclude{Unless otherwise stated}
All locks evaluated in this paper use a spin-then-park
waiting policy with a maximum spin duration of approximately 20000 cycles, where
20000 cycles is an empirically derived estimate of the average round-trip context switch time.
On SPARC the loop consists of a load and test followed by a single \texttt{RD CCR,G0}  
instruction for polite spinning.  

Broadly, we prefer that ownership of a lock passes to a more recently arrived thread.
First, threads that have waited longer are more likely to have switched from spinning to
parking, while recently arrived threads are more likely to be spinning.    
Alerting a spinning thread is cheap, relative to a thread that is fully parked. 
Threads that have waited longer are more expensive to wake, as they have less residual
cache affinity. When they run, they will incur more misses.  In addition, the 
operating system will deem that such threads have less affinity to the CPU where it 
last ran, so the scheduler, when it wakes the thread will need to make more 
complicated -- and potentially less scalable --  dispatch decisions to select the 
CPU for the thread.  The thread is more likely to migrate, and to be dispatched 
onto an idle CPU, causing costly idle-to-run transitions for that processor.

%% OPTIONAL
\subsection{Lock Handover Latency}

We define \textit{lock handover latency} as follows.   Say thread $A$ holds lock $L$ and $B$ 
waits for lock $L$.  $B$ is the next thread to acquire ownership when $A$ releases $L$.
The handover latency is the time between $A$'s call to unlock and when $B$ returns 
from lock and can enter the critical section.   Handover latency reflects
overheads required to convey ownership from $A$ to $B$.  Lock implementations 
attempt to minimize handover latency, also called \emph{responsiveness} in the literature.  
Excessive handover latency degrades scalability.
\Invisible{As noted above, if $A$ must resume $B$ via calls 
into the kernel to transition $B$ from blocked to ready, then the handover latency 
increases significantly.} 
As noted above, if $A$ must call into the kernel to wake and resume $B$, making $B$ 
eligible for dispatch, then lock handover latency increases significantly.  

\subsection{Fairness}

The default POSIX \texttt{pthread\_mutex\_lock} specification does not 
dictate fairness properties 
giving significant latitude and license to implementors. 
Fairness is considered a quality-of-implementation concern. 
In fact common mutex constructions, such as those found in Solaris or Linux, 
are based on test-and-set (TAS) locks \cite{anderson-spinning}, albeit augmented with parking, and
allow unbounded bypass with potentially indefinite starvation and unfairness.  
Similarly, the \texttt{synchronized} implementation in the HotSpot Java
Virtual Machine allows indefinite bypass as does \texttt{java.util.concurrent}
\texttt{ReentrantLock}.  
All of the above constructions ignore thread priorities for the purpose of locking.  

\Invisible{
*  This laxity also allows for CR.
*  allow; admit; affords; enables; 
*  and waiting threads fall back to parking as necessary.
*  Priorities are considered advisory; 
*  QoI = Quality of implementation 
*  QoI feature vs specification 
*  Park; block; sleep; suspend; deschedule; 
*  Liberty; License; Latitude; laxity; 
*  Dictate; Prescribe; Specify; Demand; require} 

\Invisible{
Interestingly, MCS \cite{MCS} or other strictly FIFO  locks appear 
rarely outside a few uses in operating system kernels.} 

\subsection{Succession Policies} 

Broadly, lock implementations use one of two possible \emph{succession policies}, which
describes how ownership is transferred at unlock-time when threads are waiting.  
Under \emph{direct handoff} the unlock operation passes ownership to a waiting successor, 
without releasing the lock during the transfer, enabling the successor to enter the critical section.
%% the unlock operation identifies a waiting successor and then passes ownership to that thread
If no successor exists then the lock is set to an available state.
MCS employs direct handoff.  
Under \emph{competitive succession}\cite{vm01-dice} -- also called \emph{renouncement}\cite{ipdps03-odaira} --
the owner sets the lock to an available state,
and, if there are any waiters, picks at least one as the \emph{heir presumptive},
enabling that thread to re-contend for the lock
%% OPTIONAL ...
\footnote{
The reader might note that competitive succession is analogous to the
CSMA-CD (Carrier Sense Multiple Access with Collision Detection) 
communication protocol, while direct succession is analogous to
token ring protocols.   CSMA-CD is optimistic and exhibits low latency 
under light load but suffers under high load, whereas token ring is pessimistic but 
fair, and provides stable guaranteed performance under heavy load, but incurs
more latency under light load.}
\footnote{
Competitive succession is also called \emph{barging}, as arriving threads
can barge in front of other waiting threads, allowing unbounded bypass and 
grossly unfair admission.}.
Enabling an heir presumptive is necessary to ensure progress.  
The heir presumptive may compete with arriving threads for the lock.  
TAS-based locks use competitive succession and in the simplest forms
all waiting threads act as heir presumptive and no specific enabling is needed.

\Invisible{The default Solaris and Linux pthread\_mutex implementations
allow barging, as does HotSpot Java Virtual Machine \emph{synchronized} 
construct and \texttt{java.util.concurrent} \emph{ReentrantLock}.}  

\Invisible{
*  Competition succession = Renounce; Renouncement; barging;
*  Arrange for heir presumptive
} 

\Invisible{From java.util.Concurrent AbstractQueuedSynchronizer:
Throughput and scalability are generally highest for the default barging 
(also known as greedy, renouncement, and convoy-avoidance) strategy. While this 
is not guaranteed to be fair or starvation-free, earlier queued threads are allowed 
to recontend before later queued threads, and each recontention has an unbiased 
chance to succeed against incoming threads. Also, while acquires do not ``spin'' 
in the usual sense, they may perform multiple invocations of tryAcquire interspersed 
with other computations before blocking. This gives most of the benefits of spins 
when exclusive synchronization is only briefly held, without most of the liabilities 
when it isn't. If so desired, you can augment this by preceding calls to acquire 
methods with ``fast-path'' checks, possibly prechecking hasContended() and/or 
hasQueuedThreads() to only do so if the synchronizer is likely not to be contended.} 

Locks that use direct handoff can exhibit poor performance if there are more
ready threads than CPUs and involuntary context switching -- preemption -- is in play.
The successor may have been be preempted, in which case lock handover latency will suffer.
%% and responsiveness of the lock is impaired.  
Specifically, an unlock operation
may pick thread $T$ as a successor, but $T$ has been preempted.  Circulation
stalls until the operating system eventually dispatches $T$
\footnote{Kontothanassis et al. \cite{tocs97-kontothanassis} and 
He et al. \cite{hipc05-he} suggested ways to mitigate this problem for MCS locks.}.  
This leads to the undesirable 
\emph{convoying phenomenon} \cite{sigops79-blasgen} with transitive waiting.
With competitive succession, the new owner must take explicit actions to acquire 
the lock, and is thus known to be running, albeit at just the moment of acquisition. 
Competitive succession reduces succession latency and works well in conditions
of light contention \cite{JUC-AQS}.  Direct handoff performs well under high contention 
\cite{asplos94-lim}, except when there are so many ready threads that successor 
preemption comes into play, in which case competitive succession may provide
better throughput. 

\Invisible{
*  head of line blocking 
*  Direct handoff is generally better under high contention while competitive 
   succession is more optimistic and reduces succession latency in conditions 
   of light contention.  
*  Three modes :
   Light or no contention : use competitive succession
   heavy contention : use direct handoff
   Involuntary preemption : use competitive succession
} 

Direct handoff suffers from an additional performance concern related
to the waiting policy.  If the successor $T$ parked itself by calling into the 
operating system, then the unlock operator needs to make a corresponding system call 
to wake and unpark $T$, making $T$ eligible for dispatch.  The time from an unpark($T$)
call until the corresponding blocked thread $T$ returns and resumes from park can 
be considerable. Latencies of more than 30000 cycles are common even in the 
best case on an otherwise unloaded system where there are fewer ready threads than 
CPUs and an idle CPU is available on which to dispatch $T$ 
\footnote{Unpark itself incurs a cost of more than 9000 cycles to the caller on our SPARC T5 system.}.
\Invisible{We have observed latencies in excess on 80000 cycles on modern Intel processors
when unpark causes a thread to be dispatched onto a CPU idling in a deep low-power sleep state.} 
Crucially, these administrative latencies required by succession to resume threads accrue while 
the lock is held, artificially lengthening the critical section.  Such lock handover 
latency greatly impacts throughput over the contented lock, and can dominate 
performance under contention.  
Direct handoff is generally not preferred for locks that wait 
via parking as context switch overheads artificially increase the critical section 
duration and effective lock hold times.   

\Invisible{
*  Not preferred vs generally unsuitable
*  Blocks and resumes vs park-unpark 
*  The most recently arrived threads are the most likely to still be 
   spinning, but they will be the last to be granted the lock.} 

%% REDUNDANT with above ...
\Invisible{
Lock algorithms can provide succession either by direct handoff -- where
ownership of the lock is conveyed directly from the current owner to some
waiting thread -- or via so-called competitive succession, where the
current owner, in unlock(), releases the lock and allows waiting threads
to contend for the lock.  Direct handoff is generally better under high
contention while competitive succession is more optimistic and reduces
succession latency in conditions of light contention.  To provide progress
and liveness, locks that use competitive succession may need to unpark an 
``heir presumptive'' thread that had been waiting.  The heir presumptive
can then compete for the lock.  } 

\Invisible{
Direct handoff typically implies the existence of an explicit list of waiting threads.  
In turn, that allows local spinning by those waiters.  
This claim is not universally correct -- ticket locks serve as counter-example.} 

All strictly FIFO locks use direct handoff.  Relatedly, all locks that 
use \emph{local spinning} \cite{topc15-dice}, such as MCS, also use direct handoff.  
With local spinning, at most one waiting thread spins on a given location at any given time.
Local spinning often implies the existence of an explicit list of waiting threads 
\footnote{More precisely, at unlock-time the owner thread must be able to identify 
the next waiting thread -- the successor.}. 
Depending on the platform, local spinning may reduce the ``invalidation diameter'' 
of the writes that transfer ownership, as the location to be written should be monitored
by only one thread and thus reside in only one remote cache.
Lock algorithms such as TAS use \emph{global spinning}, where all threads waiting
on a given lock busy-wait on a single memory location  
\footnote{Systems with MOESI-based cache coherence may be more tolerant of
global spinning than those that use MESI \cite{topc15-dice}.}. 
\Invisible{Invalidation diameter; number of participating caches;} 

%% OPTIONAL 
Given its point-to-point nature where thread $A$ directly unparks and wakes $B$, 
using park-unpark for locks requires the lock algorithm to maintain an 
explicit list of waiting threads, visible to the unlock operator \cite{blog-waitingpolicies-stp}.  
Most locks that use local spinning, such as MCS, can therefore be
readily converted to use parking.  
A simple TAS lock with global spinning and competitive succession requires 
no such list be maintained -- the set of of waiting threads is implicit and 
invisible to the unlock operator.  
Lock algorithms that use global spinning, such as ticket locks or TAS locks, 
are more difficult to adapt to parking.  
%% OPTIONAL ...
As noted above, parking is typically 
inimical to locks that use direct handoff, as the context switch overheads 
artificially increase the critical section length.

We note the following tension. Locks, such as MCS, that use succession by direct 
handoff and local spinning can be more readily adapted to use spin-then-park waiting,
the preferred waiting policy. 
Under high load, however, with long waiting periods, direct handoff can interact 
poorly with parking because of increased handover latency, where the successor has reverted 
to parking and needs to be explicitly made ready.  Spinning becomes less successful
and the lock devolves to a mode where all waiting threads park. 
MCSCR uses direct handoff, but can provide relief, relative to a pure FIFO lock, from 
handover latency as the successor is more likely to be spinning instead of fully 
parked.

%% Use struts to increase vertical white-space in table cells ...
\newcommand\TopStrut{\rule{0pt}{2.6ex}}       
\newcommand\BottomStrut{\rule[-1.2ex]{0pt}{0pt}} 
%% Alternatives : 
%% *  \setlength\extrarowheight{3pt} from array package
%% *  \bigstrut
%% *  \hline \noalign{\vskip 2mm} 
%% *  \addlinespace[2ex]

\begin{figure*}[h]
%% \begin{center}
\begin{tabular}{|l|c|c|}
\hline
\TopStrut \BottomStrut
Property : Lock & TAS & MCS  \\
\hline
Succession & Competitive & Direct \\
Able to use spin-then-park waiting & No & Yes \\
Uses ``polite'' local spinning to minimize coherence traffic & No & Yes \\
Low contention performance -- Latency & Preferred & Inferior to TAS \\
High contention performance -- Throughput & Inferior to MCS & Preferred \\ 
Performance under preemption & Preferred & Suffers from lock-waiter preemption \\
Fairness & Unbounded unfairness via barging & Fair \\
Requires back-off tuning and parameter selection & Yes & No \\
\hline
\end{tabular}
%% \end{center}
\caption{Comparison of TAS and MCS locks}
\label{Table:TASvsMCS}
\end{figure*}

%% \section{Experimental Results} 
%% \section{Performance}
%% \section{Lock Implementations}

\section{Evaluation} 

We used an Oracle SPARC T5-2 \cite{T5-2} for all experiments.  The T5-2 has 2 sockets,
each with a single T5 processor running at 3.6 GHz.  Each processor has 16 cores, and each core has
2 pipelines supporting 8 logical CPUs (``strands''), yielding 128 logical CPUs per socket.  
%% REDUNDANT: see "fuse" above
If there is only one active CPU on a core, both pipelines promptly and automatically
fuse to provide improved performance.  
%% OPTIONAL ...
The extra strands exist to exploit available memory-level parallelism (MLP) \cite{isca04-chou}.  
%% Exploit; leverage; utilize
Each socket has an 8MB unified L3 LLC shared by all cores on that socket.  Each core has
a fully associative 128-entry data TLB shared by all logical CPUs on that core. 
Each TLB entry can support all the available page sizes.  Each core also has 
a 16KB L1 data cache and a 128KB L2 unified cache.  
\Exclude{The L1 uses a write-back  policy while the L2 and L3 use a write-through policy.  
The system uses a MOESI cache coherence protocol.}  
%% cache coherence protocol vs cache coherence states 
For all experiments we took all the CPUs on the second T5-2 socket offline,
yielding a non-NUMA T5 system with 128 logical CPUs.  
All data collected for this paper was run in maximum performance mode with power
management disabled.  
%% OPTIONAL ...
The SPARC T5 processor exposes the \texttt{sel\_0\_ready} hardware performance counter which
tallies the number of cycles where logical CPUs were ready to run, but pipelines where
not available.  This counter is used to detect and measure pipeline oversubscription and
competition.   

The system ran Solaris 5.11. 
Unless otherwise specified, all code was compiled with gcc 4.9.1 in 32-bit mode.
We observed that the performance and scalability of numerous benchmarks were sensitive to
the quality of the \texttt{malloc-free} allocator.  
The default Solaris allocator protects the heap with a single global lock and scales poorly.  
The poor performance of the default allocator often dominated overall performance of applications,
and masked any sensitivity to lock algorithms.  We therefore used the scalable LD\_PRELOAD \emph{CIA-Malloc} 
allocator \cite{ISMM11-Afek-CIA} for all experiments, except where noted.  CIA-Malloc does not itself
use the \texttt{pthread\_mutex} primitives for synchronization.  

All locks were implemented as LD\_PRELOAD interposition libraries, exposing
the standard POSIX \texttt{pthread\_mutex} programming interface.  
\Exclude{The pthreads interface gives the implementor considerable latitude and license 
as to admission schedule and fairness.} 
LD\_PRELOAD interposition allows us 
to change lock implementations by varying the LD\_PRELOAD environment variable and 
without modifying the application code that uses locks.  

We use the default free-range threading model, where the operating system is free to migrate 
threads between processor and nodes in order to balance load or achieve
other scheduling goals.  Modern operating systems use aggressive intra-node migration to 
balance and disperse the set of ready threads equally over the 
available cores and pipelines, avoiding situations where some pipelines are 
overutilized and others underutilized \footnote{We observe that explicit binding
of threads to CPUs or indefinite spinning precludes this benefit.}. 
Inter-node migration is relatively expensive and is less frequent.  
\Invisible{equally, equitably, uniformly} 

We use a number of small carefully constructed benchmarks to exhibit
various modes of contention for shared hardware resources.  The first examples
are intentionally simple so as to be amenable to analysis. 
\Invisible{Exemplars; contrived to illustrate; designed to show existence of effect;} 

%% OPTIONAL : relegate Gini to footnote or expunge ...
%% Variations : describes; informs us of; 
We measure long-term fairness with the relative standard deviation (RSTDDEV), which
describes the distribution of work completed by the set participating threads.  We also
report the Gini Coefficient \cite{gini,blog-longtermfairness}, 
%% phrase : by|over the set of participating threads
%% a statistic which describes the distribution of work completed over the set of threads.   
%% The Gini Coefficient -- which is equal to half the relative mean absolute difference -- 
popular in the field of economics as in index of income disparity and unfairness.  
A value of $0$ is ideally fair (FIFO), and $1$ is maximally unfair.  
%% A FIFO MCS lock \cite{MCS}, for instance, would be expected to yield a Gini Coefficient near $0$.  
%% We also report the relative standard deviation (RSTDDEV) of
%% the distribution of work completed by the participating threads. 

\Exclude{Reporting long-term and short-term fairness metrics requires 
modifications to the application.  As such, we only collect and report values 
for our own synthetic microbenchmarks. } 

%% Possibly make the following a footnote ...
\Invisible{We might also the compute the Gini 
Coefficient over smaller intervals in the acquisition history and then combine
those values to form an index that represents short-term fairness.  We believe, 
however, that LWSS is more intuitive as a measure for short-term fairness as the value is expressed 
in units of threads.}  

\Exclude{Consider weighted metrics that combine fairness and aggregate 
throughput into a composite score : 
$AggregateTput * (1-Gini)$ ; 
$MinTput * N$ ; 
$AggregateTput * (AveragePut/MaxTput)$  
MinTPut is the tput of the slowest thread.
See exbench-stdmap.c.}

\subsection{Random Access Array}

\begin{figure}[ht] 
\vspace{-0pt} 
\includegraphics[rviewport={.1 0 1 1},clip,width=6.4cm,angle=270,origin=c]{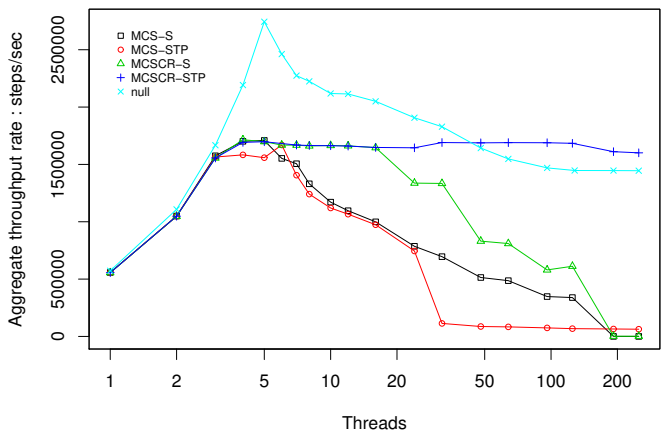}
\vspace{-60pt}  %% tighten vertical spacing
\caption{Random Access Array} 
\label{Figure:randarray} 
\end{figure}

The \texttt{RandArray} microbenchmark spawns $N$ concurrent threads.  Each thread loops as follows:
acquire a central lock $L$; execute a critical section (CS); release $L$; 
execute a non-critical section (NCS).  At the end of a 10 second measurement
interval the benchmark reports the total number of aggregate iterations completed by all
the threads.  RandArray also reports average LWSS, median time to reacquire, and long-term fairness statistics.  
%% such as the Gini Coefficient, on the distribution of iterations completed over the set of threads. 
%% OPTIONAL begin ...
\Exclude{
The modifications to report the LWSS and median time to reaquire were done in such a manner so as to
minimize the probe effect.  Each thread maintains a local acquisition history.  The benchmark
merges those histories after the measurement interval to form a global history, from 
which the average LWSS and median time to reacquire are computed. } 
%% OPTIONAL end 
We vary $N$ and the lock algorithm and report aggregate throughput results in 
Figure \ref{Figure:randarray}, taking the median of 7 runs.  The number of threads 
on the X-axis is shown in log scale. 

The NCS consists of an inner loop of 400 iterations.  Each iteration generates a 
uniformly distributed random index into a thread-private array of 256K 32-bit integers, 
and then fetches that value.  To avoid the confounding effects of coherence traffic, 
we used only loads and no stores.  The CS executes the same code, but has a duration of 100
iterations and accesses a shared array of 256K 32-bit integers.  
The ideal speedup is 5x.  
The 1MB arrays reside on large pages to avoid DTLB concerns. 
%% OPTIONAL ...
The random number generators are thread-local.  
%% OPTIONAL ...
We used random indexes to avoid the impact of automatic hardware prefetch mechanisms
\footnote{Our benchmark was inspired by ``new benchmark'' from \cite{HBO}}. 

\texttt{MCS-S} is the classic MCS algorithm where the waiting loop is augmented to include
a polite \texttt{RD CCR,G0} instruction.  \texttt{MCS-STP} uses spin-then-park waiting.
\texttt{MCSCR-S} is MCSCR where the waiting loop uses the \texttt{RD CCR,G0} instruction
on every iteration, and \texttt{MCSCR-STP} is MCSCR with spin-then-park waiting.  
For reference, we include \texttt{null} where the lock acquire and release operators
are degenerate and return immediately.  
\texttt{Null} is suitable only for trivial microbenchmarks, as other more 
sophisticated applications will immediately fail with this lock.

As we can see in Figure \ref{Figure:randarray}, ignoring \texttt{null}, the \emph{peak} 
appears at about $N=5$, where the maximum observed speedup is slightly more than 3 times
that of a single thread.  MCS-S and MCS-STP start to show evidence of collapse at 
6 threads where the total NCS and CS footprint is 7MB, just short of the total 
8MB LLC capacity.  The LLC is not perfectly associative, so the onset of thrashing
appears at footprints slightly below 8MB.  Absent CR, the NCS instances erode
LLC CS residency and impair scalability.  As noted above, MCS-STP performs poorly
because spin-then-parking waiting is unsuitable for direct handoff FIFO locks such
as MCS.  Crucially, spin-then-park delivers good performance for MCSCR over all thread
counts, but decreases performance of classic MCS except in the case where there are more
ready threads than CPUs, where pure unbounded spinning breaks down.   
Interestingly, MCSCR-STP achieves better performance than \texttt{null} beyond 48 threads. 

While not immediately visible in the figure, at 256 threads MCS-STP yields 120x better 
throughput than MCS-S.  Under MCS-S, as we increase the number of ready spinning 
threads, we increase the odds that the lock will be transferred to a preempted successor, 
degrading performance.  Spinning threads must exhaust their allotted time slice until the
owner is eventually scheduled onto a CPU.   At 256 threads, MCS-STP requires a voluntary
context switch for each lock 
handover, but it sustains reliable and consistent -- but relatively low -- performance even if we 
further increase the number of threads.  This demonstrates why lock designers
conservatively opt for parking over unbounded spinning.  Typical time slice periods used
by modern operating systems are far longer than park-unpark latencies.  As such,
we prefer progress via voluntary context switching over involuntary preemption.  

To confirm our claim of destructive interference and thrashing in the LLC, we implemented
a special version of \texttt{RandArray} where we modeled the cache hierarchy of the system
with a faithful functional software emulation, allowing us to discriminate 
\emph{intrinsic self-misses}, where misses are caused by a CPU displacing lines that
it inserted, versus \emph{extrinisc} misses caused by sharing of a cache.  
We augmented the cache lines in the emulation with a field that identified which
CPU had installed the line.  That we know of, no commercially available CPU design
provides performance counters that allow misses to be distinguished in this manner,
although we believe such a facility would be useful.  
All data in this paper is derived from normal runs without the emulation layer.  
\Invisible{Self-vs-self and Self-vs-other displacement} 

\Invisible{
*  Time Slice; quanta;  
*  Burn through; exhaust; deplete; consume; complete; execute to the end of;
*  Favor; Prefer; 
*  if forced to choose ...
} 

In addition to competition for LLC residency, this graph reflects competition for
pipelines \footnote{Other core-level resources such as TLB residency are similarly
vulnerable to competition and can benefit from CR.}.  
At 16 threads -- recall that we have 16 cores -- we see MCSCR-S fade.
In this case the spinning threads in the PS compete for pipelines with the ``working''
threads in the ACS.  (The polite spin loop helps reduce the impact of pipeline 
competition, which would otherwise be far worse).  
Using a spin-then-park waiting strategy avoids this concern.
MCSCR-STP manages to avoid collapse from pipeline competition. 

MCS-S and MCS-STP depart from MCSCR-S and MCSCR-STP at around 8 threads because
of LLC thrashing.  MCSCR-S departs from MCSCR-STP at 16 threads because of competition
for pipelines.  The slow-down arises from the spin-only waiting policy of those locks.
MCS-S and MCSCR-S exhibit an abrupt cliff at 128 threads because
of competition for logical CPU residency arising from unbounded spinning.
Beyond 128 threads there is system-wide competition for logical processors.
MCSCR-STP is the only algorithm that maintains performance in this region,
again reflecting the importance of waiting policies.

In Figure \ref{Table:randarray-details} we include more details of RandArray execution
at 32 threads.  The L3 miss rate is considerably lower under the CR forms.  
As would be expected, the average LWSS and the CPU utilization correspond closely under
MCSCR-STP. Note too that the CPU utilization for MCSCR-STP is low,
providing lower energy utilization and improved opportunities for multi-tenancy.  Despite
consuming the least CPU-time, MCSCR-STP yields the best performance. 
We use the Solaris \texttt{ldmpower} facility to measure the wattage above idle, showing
that power consumption is also the lowest with MCSCR-STP.  
As evidenced by the LWSS and MTTR values, CR-based locks reduce the number of distinct NCS
instances accessed in short intervals, in turn reducing pressure and miss rates in the LLC,
accelerating CS execution, and improving overall throughput.

%% Issue : prefer figures all placed on one page  
%% keywords : Group; place; placement; cluster; affinity; proximity
%% To force the figures to be grouped on one page :
%% +  Minipages
%% +  \subfloat
%% +  Use just a single "figure" environment with multiple elements inside
%% +  \includepackage{placeins} then \FloatBarrier around figure blocks
%% +  subfigure
%% +  Use \begin{figure}[h] : h = "here" advisory hint
%%    !htp; H vs h; !Htp vs tp

%% Issue : Excess Vertical Whitespace in graphics produced by R
%% keywords : Clip; trim; crop
%% use vspace{-20pt} to tighten vertical spacing 
%% Variations : 
%% \includegraphics[rviewport={.1 0 1 1},clip,width=6.4cm,angle=270,origin=c]{plot.eps}
%% \includegraphics[width=7cm,angle=270,origin=c,trim={0 0 0 7cm}]{plot.eps}
%% \includegraphics[width=5.8cm,angle=270,origin=c,trim={3cm 0 0 1.2cm},clip]{plot.eps}
%% The above seem to work 
%% \epsfxsize=2.20in \epsfbox{misses.eps} \resizebox{...} 
%% \includegraphics[angle=270,origin=c,width=16cm]{plot-avl/plot.eps}

%% Consider also : CPI
%% Beware the CPI and L3-misses-Per-1K instructions are influenced by 
%% the waiting policy.  Busy-waiting instructions count as instructions.  
%% use either footnotesize or scriptsize
%% \fontsize{7}{7}\selectfont
%% \scalebox{.7}{....} 

\begin{figure}[h]
{ \fontsize{6.5}{6.5}\selectfont
\begin{center}
\begin{tabular}{|l|c|c|c|c|}
\hline
\TopStrut \BottomStrut
Locks & MCS-S & MCS-STP & MCSCR-S & MCSCR-STP \\
\hline
\hline
\TopStrut \BottomStrut
Throughput                      (ops/sec) & 0.7M   & 0.1M   & 1.3M  & 1.6M  \\
Average LWSS                    (threads) & 32     & 32     & 5.3   & 5.1   \\
MTTR                            (threads) & 31     & 31     & 3     & 3     \\
Gini Coefficient                          & 0.001  & 0.001  & 0.076 & 0.078 \\ 
RSTDDEV                                   & 0.000  & 0.000  & 0.152 & 0.155 \\
Voluntary Context Switches                & 0      & 798K   & 11    & 6K    \\
CPU Utilization                           & 32x    & 16.8x  & 32x   & 5.2x  \\
L3 Misses                                 & 11M    & 10M    & 152K  & 172K  \\
$\Delta$ Watts above idle                 & 113    & 79     & 91    & 63    \\
\hline
\end{tabular}
\end{center}
%% \vspace{-0.1in}
\caption{In-depth measurements for Random Access Array benchmark at 32 threads and
a 10 second measurement interval} 
\label{Table:randarray-details}
\vspace{-0.1in}
}
\end{figure}

\begin{figure}[h]
\vspace{-10pt}  %% tighten vertical spacing
\includegraphics[rviewport={.1 0 1 1},clip,,width=6.4cm,angle=270,origin=c]{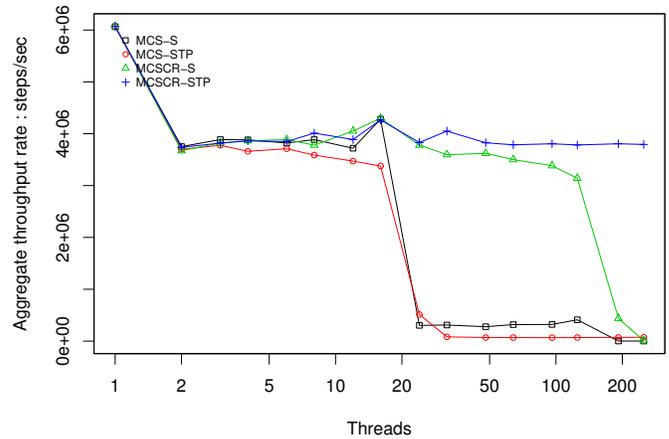}
\vspace{-60pt}  %% tighten vertical spacing
\caption{Core-level DTLB Pressure} 
\label{Figure:dtlb} 
\end{figure}

\begin{figure}[h]
%% \vspace{-10pt} 
\includegraphics[rviewport={.1 0 1 1},clip,,width=6.4cm,angle=270,origin=c]{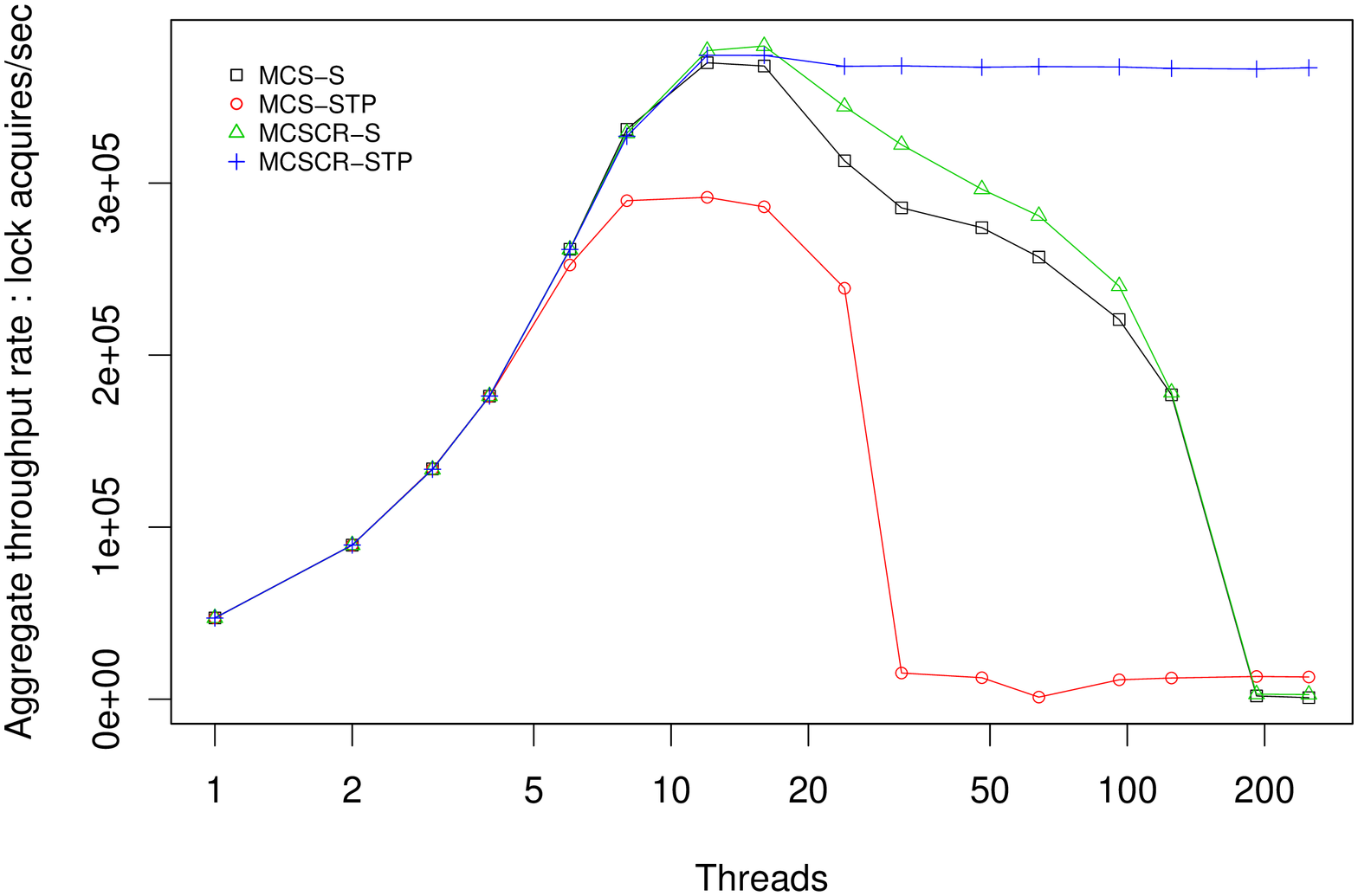}
\vspace{-60pt}  %% tighten vertical spacing
\caption{libslock} 
\label{Figure:libslock} 
\end{figure}

\begin{figure}[h] 
\vspace{-10pt}  %% tighten vertical spacing
\includegraphics[rviewport={.1 0 1 1},clip,,width=6.4cm,angle=270,origin=c]{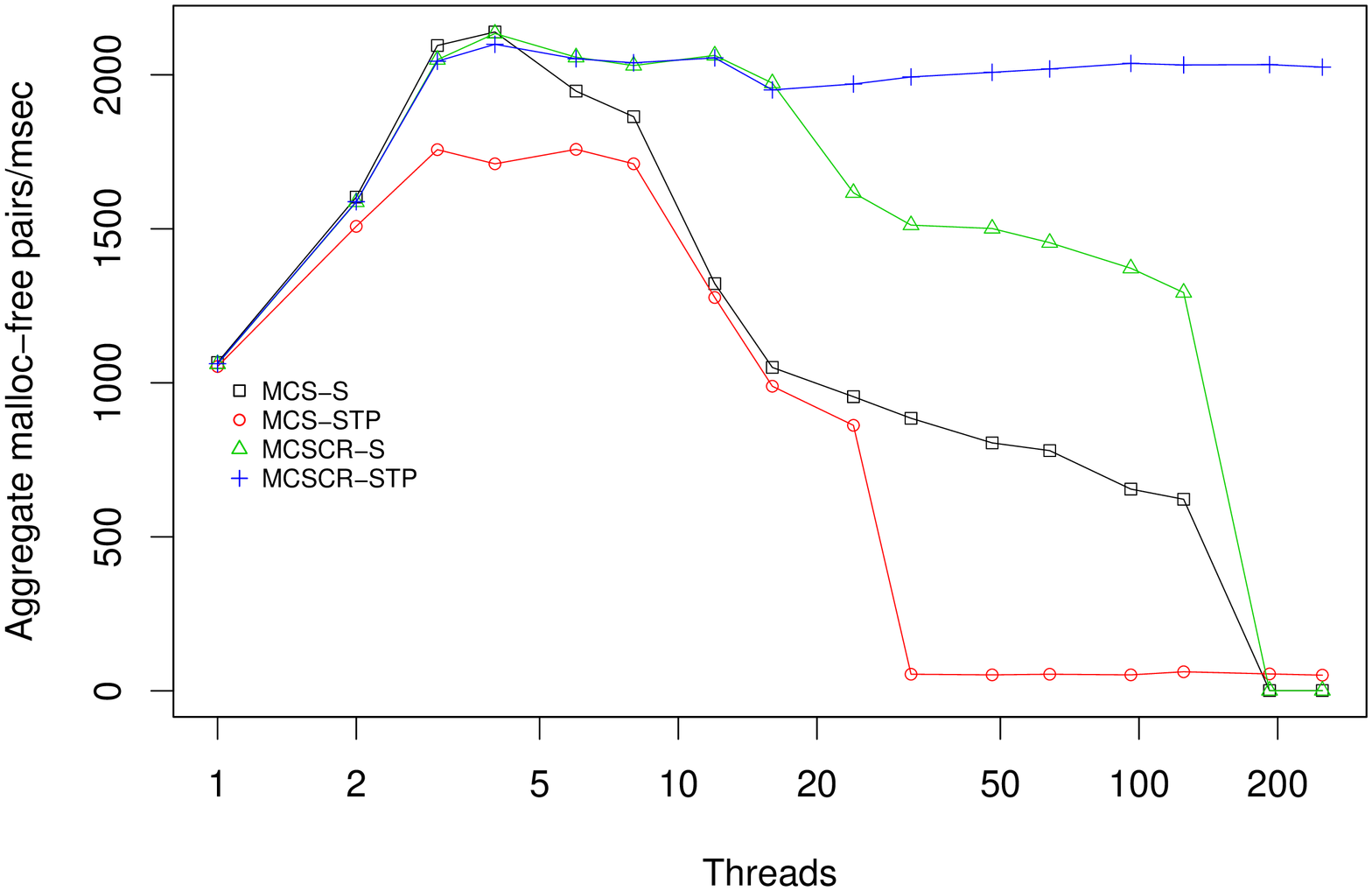}
\vspace{-60pt}  %% tighten vertical spacing
\caption{mmicro} 
\label{Figure:mmicro} 
\end{figure}

\begin{figure}[h] 
\vspace{-10pt}  %% tighten vertical spacing
\includegraphics[rviewport={.1 0 1 1},clip,,width=6.4cm,angle=270,origin=c]{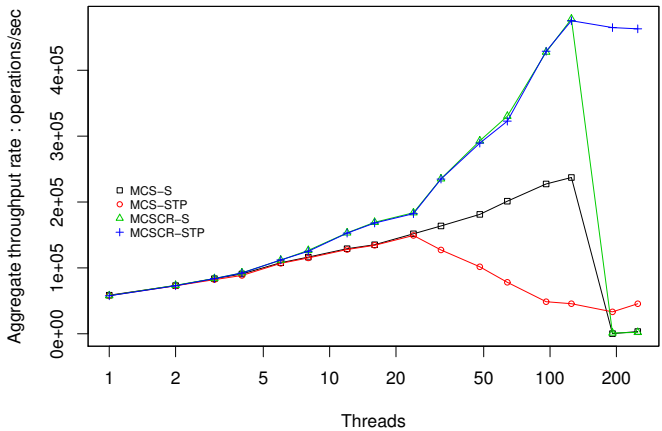}
\vspace{-60pt}  %% tighten vertical spacing
\caption{leveldb readwhilewriting benchmark} 
\label{Figure:leveldb} 
\end{figure}

\begin{figure}[h]
\vspace{-10pt}  %% tighten vertical spacing
\includegraphics[rviewport={.1 0 1 1},clip,,width=6.4cm,angle=270,origin=c]{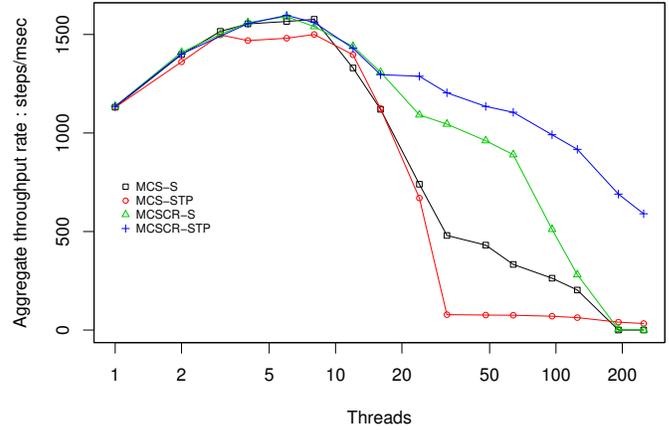}
\vspace{-60pt}  %% tighten vertical spacing
\caption{KyotoCabinet kccachetest} 
\label{Figure:kccachetest} 
\end{figure}

\begin{figure}[h] 
\vspace{-10pt}  %% tighten vertical spacing
\includegraphics[rviewport={.1 0 1 1},clip,,width=6.4cm,angle=270,origin=c]{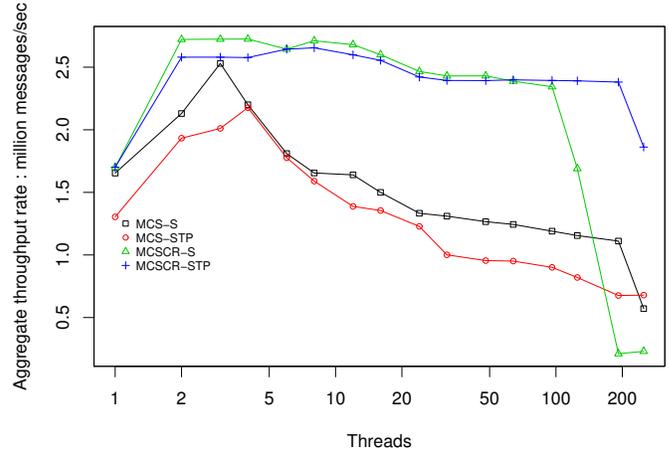}
\vspace{-60pt}  %% tighten vertical spacing
\caption{producer\_consumer with 3 consumer threads} 
\label{Figure:producerconsumer} 
\end{figure}

\begin{figure}[h] 
\vspace{-10pt}  %% tighten vertical spacing
\includegraphics[rviewport={.1 0 1 1},clip,,width=6.4cm,angle=270,origin=c]{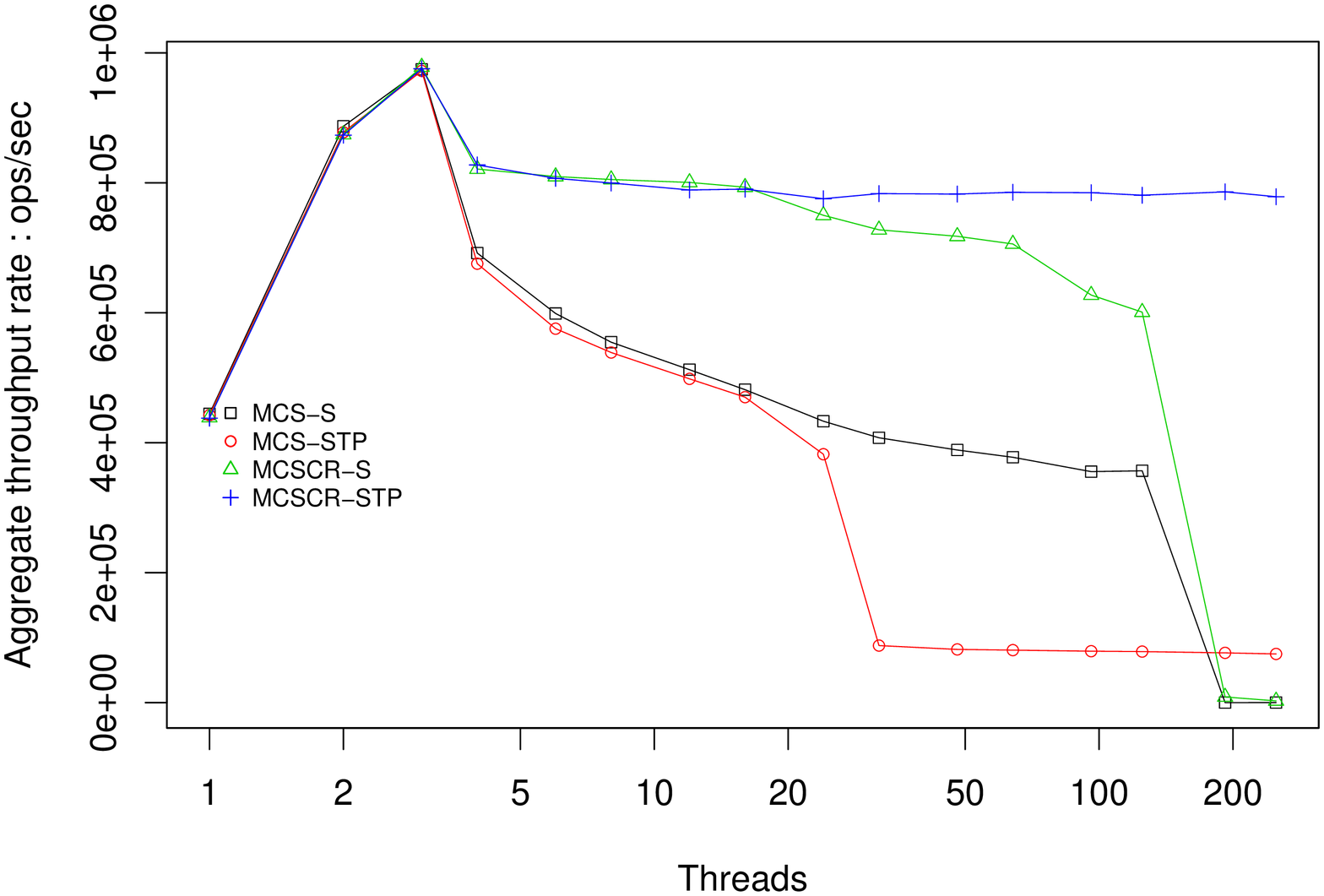}
\vspace{-60pt}  %% tighten vertical spacing
\caption{keymap} 
\label{Figure:keymap} 
\end{figure}

\begin{figure}[h] 
\vspace{-10pt}  %% tighten vertical spacing
\includegraphics[rviewport={.1 0 1 1},clip,,width=6.4cm,angle=270,origin=c]{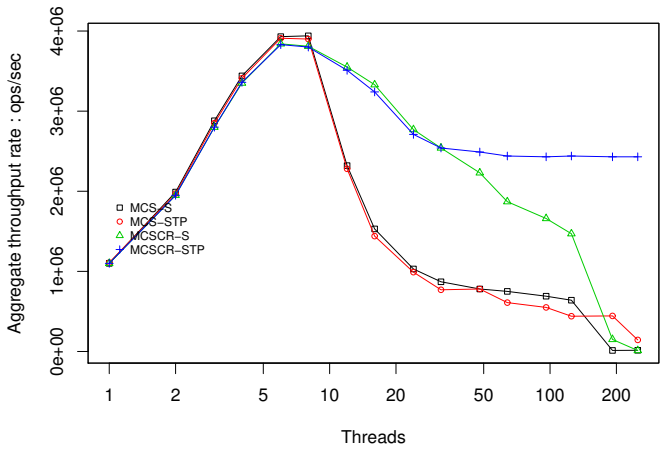}
\vspace{-60pt}  %% tighten vertical spacing
\caption{LRUCache} 
\label{Figure:LRUCache} 
\end{figure}

\begin{figure}[h] 
\vspace{-10pt}  %% tighten vertical spacing
\includegraphics[rviewport={.1 0 1 1},clip,,width=6.4cm,angle=270,origin=c]{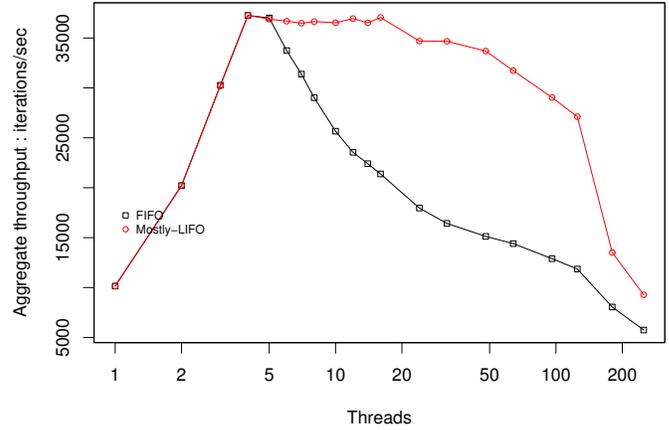}
\vspace{-60pt}  %% tighten vertical spacing
\caption{RandArray transliterated to perl} 
\label{Figure:perl} 
\end{figure}

\begin{figure}[h] 
\vspace{-10pt}  %% tighten vertical spacing
\includegraphics[rviewport={.1 0 1 1},clip,,width=6.4cm,angle=270,origin=c]{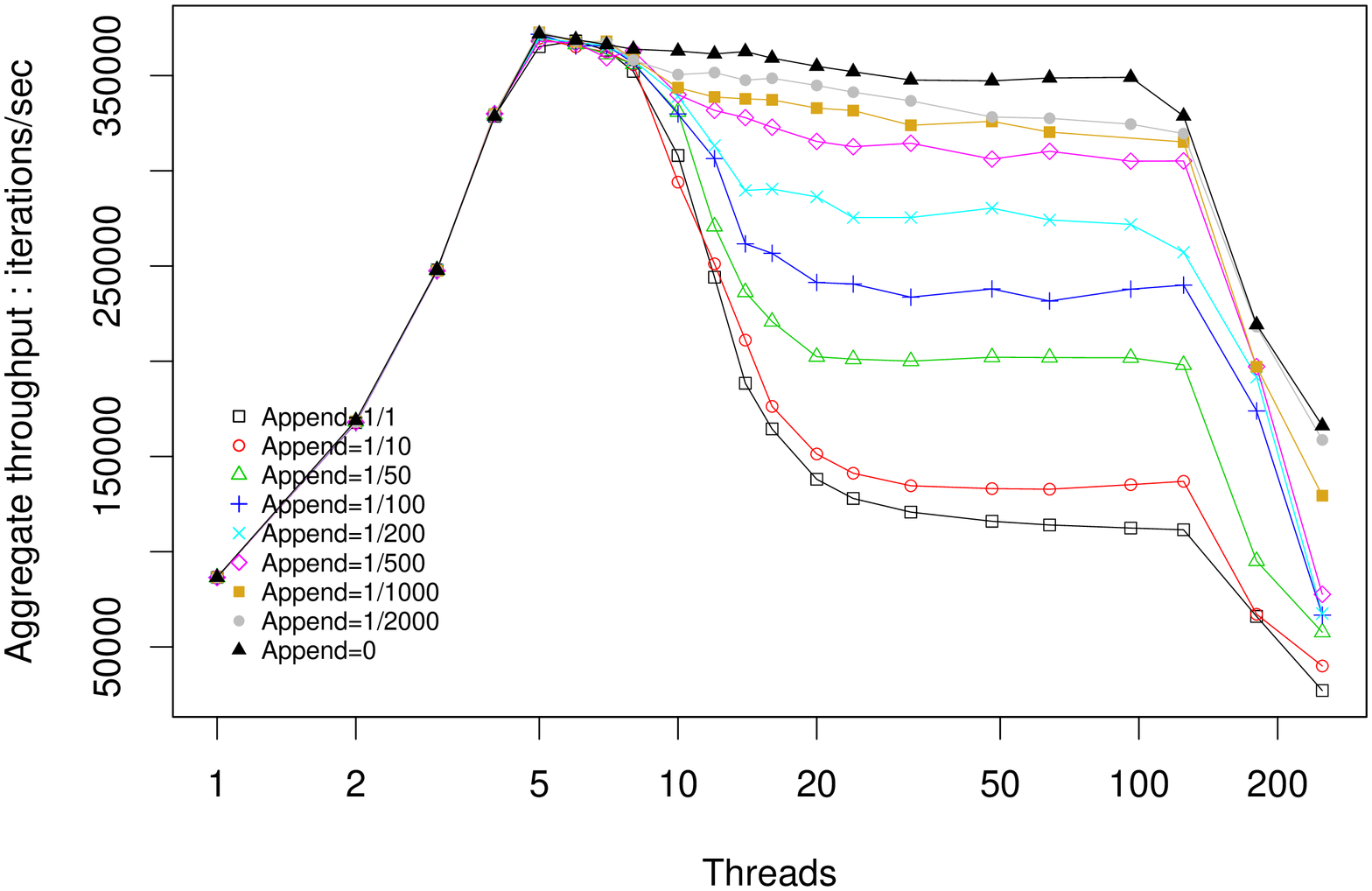}
\vspace{-60pt}  %% tighten vertical spacing
\caption{Buffer Pool} 
\label{Figure:BP} 
\end{figure}

%% Consider: trim,clip

\subsection{DTLB Pressure} 

%% phrasing: Demonstrated an example of ...

Figure \ref{Figure:randarray} demonstrated competition for
socket-level LLC.  In Figure \ref{Figure:dtlb} we now provide an illustration of 
core-level DTLB pressure.  
The structure of our \texttt{RingWalker} benchmark is similar to that of 
\texttt{RandArray}.  Each thread has a private circularly linked list.  Each list element
is 8KB in length and resides on its own page.  Each ring contains 50 elements.  
The non-critical section iterates over 50 thread-private elements.  We record the last element
at the end of the NCS and start the next NCS at the subsequent element.  The critical
section is similar, although the ring is shared, and each critical section advances
only 10 elements.  The inflection point at 16 threads for MCS-S and MCS-STP is attributable 
to DTLB misses.  Recall that each SPARC core has 128 TLB entries.  When two members of the ACS
reside on the same core, we have a total DTLB span of 150 pages, which exceeds the
number of TLB entries.  (The CS ring has a span of 50 pages, and each of the 2 
NCS instances have a span of 50 pages).  We can shift the inflection point for MCS-S and
MCS-STP to the right by decreasing the number of elements in the rings. 
The cache footprint of a ring with $N$ elements is just $N$ cache lines, and
the DTLB footprint is $N$ entries. 
The offsets of elements within their respective pages were randomly
colored to avoid cache index conflicts. 

\subsection{libslock} 

Figure \ref{Figure:libslock} shows the performance of the \texttt{stress\_latency} benchmark
from \cite{SOSP13-David} \footnote{We use the following command line: 
\texttt{./stress\_latency -l 1 -d 10000 -a 200 -n <threads> -w 1 -c 1 -p 5000}.}. 
The benchmark spawns the specified number of threads, which all run concurrently
during a 10 second measurement interval.  Each thread iterates as follows:
acquire a central lock; execute 200 loops of a delay loop; release the lock; 
execute 5000 iterations of the same delay loop.  The benchmark reports 
the total number of iterations of the outer loop. 
This delay loop and thus the benchmark itself are cycle-bound, and the main inflection
point appears 16 threads where
threads that wait via spinning compete with working threads for core-level pipelines.
This again demonstrates the impact of waiting policy.  
Similar to many other synthetic lock microbenchmarks, very few distinct locations are accessed: 
there is only one shared variable and there are no memory accesses within the 
non-critical section.  

\Invisible{In practice, lock-based code would be expected to display
more variety.  A lock algorithm would be expected to perform well on such 
a simple benchmark, but the benchmark is not likely to be reflective of
common usage.} 

\subsection{malloc scalability benchmarks}

In Figure \ref{Figure:mmicro} we use the \texttt{mmicro} malloc-free scalability
benchmark from \cite{topc15-dice}.  In this case we use the default Solaris \texttt{libc}
memory allocator, which is implemented as a splay tree protected by a central mutex.  
While not scalable, this allocator yields a dense heap and small footprint and
thus remains the default. 
\texttt{Mmicro} spawns a set of concurrent threads, each of which iterates as follows: 
allocate and zero 1000 blocks of length 1000 bytes and then release those 1000 blocks.  The 
measurement interval is 50 seconds and we report the median of 7 runs.  The benchmark 
reports the aggregate malloc-free rate.  Each malloc and free operation acquires the 
central mutex.  The benchmark suffers from competition for LLC residency, and, at above 
16 threads, from pipeline competition.  
Under CR, fewer threads circulating over the central mutex lock in a given period also yields fewer 
malloc-ed blocks in circulation which in turn yields better hit rates for caches and core-level 
DTLBs.

\Invisible{ 
%% \section{malloc} 
%% malloc-solaris; mmicro; runset-malloc

In addition to the usual benefits of CR noted above, CR provides 
additional and surprising modes of benefits for simple malloc-free allocators.
Assume we have an allocator that protects a unified heap with a single
\texttt{pthread\_mutex}.  The default Solaris \texttt{libc} allocator is of this design.  
Our application, \texttt{mmicro},  spawns $N$ concurrent threads that each loop,
allocating a set of blocks via \texttt{malloc} and then releasing those blocks.  
Furthermore assume the malloc lock is highly contented and that we are using
an LD\_PRELOAD interposition library to replace the normal \texttt{pthread\_mutex}
with a lock providing CR.  As such we will enjoy fewer threads circulating
over the lock in unit-time.  Fewer threads circulating implies that fewer malloc-ed 
blocks are circulating.  In turn, this yields better hits rates in the core-level
DTLBs and caches.  Consider an ``episode'' -- a sequence of malloc
operations performed by a given thread in the \texttt{mmicro} benchmark.  
CR provides the following benefits:
\begin{itemize}
\item inter-episode L1-DC locality and reuse
\begin{itemize} 
\item Reduces both normal capacity misses 
\item Reduces write invalidation -- coherence misses
\end{itemize}
\item inter-episode DTLB locality and reuse
\item intra-episode DTLB span ; TLB footprint of set of blocks in episode ; working set
\end{itemize} 
}
 
\subsection{leveldb benchmark} 

In Figure \ref{Figure:leveldb} we used the \texttt{db\_bench readwhilewriting} 
benchmark in \texttt{leveldb} version 1.18 database\footnote{\url{leveldb.org}}, varying 
the number of threads and reporting throughput from the median of 5 runs.  
Both the central database lock and internal \texttt{LRUCache} locks are 
highly contended, and amenable to CR, which reduces the last-level cache miss rate.  
We used the default Solaris malloc-free allocator for this experiment.  

\subsection{Kyoto Cabinet kccachetest}

%% The benchmark uses POSIX reader-writer locks, and performance is
%% known to be sensitive to the lock algorithm \cite{ppopp13-calciu}.  
%% We modified it to use POSIX \texttt{pthread\_mutex} locks

In Figure \ref{Figure:kccachetest} we show the benefits of CR for the Kyoto Cabinet \cite{kyotocabinet} 
\texttt{kccachetest} benchmark, which exercises an in-memory database.
The performance of the database is known to be sensitive to the choice of lock algorithm
\cite{ppopp13-calciu}.  
We modified the benchmark to use standard POSIX \texttt{pthread\_mutex} locks 
%% OPTIONAL : 
\Exclude{
\footnote{Normally it is unsafe to convert
reader-writer locks to mutexes, as the program may reasonably expect reader-reader
concurrency.  Threads $A$ and $B$ might acquire read permission and then expect
to communicate with each in a lock-step synchronous fashion.  This expectation
would be violated if we converted reader-writer locks to simple mutexes.  But in 
the specific case of \texttt{kccachetest} we can safely adapt the benchmark to 
uses mutexes.}}  
and to run for a fixed time and then report the aggregate work completed.  
We used a 300 second measurement interval and took the median of 3 runs.  Finally, 
the key range for a run was originally computed as a function of the number of threads, 
making it difficult to compare scaling performance while varying the thread count.  
We fixed the key range at 10M elements.  
%% Make graphs problematic
%% FWRT vs FTRW

Peak performance occurs at 5 threads, dropping rapidly as we increase
the number of threads.  Analysis of the program with hardware performance counters
%% shows the LLC miss rate is responsible for the drop between 5 and 16 threads.
shows a marked increase in LLC miss rate above 5 threads.
After 16 threads MCS-S and MCS-STP suffer from both increasing LLC misses and
from pipeline competition.   MCSCR-STP manages to avoid the collapse exhibited
by the basic MCS forms.  

\subsection{producer-consumer benchmark}

%% OPTIONAL : producer_consumer has a different and less general mode-of-benefit.
%% MoB for other examples involves competition for shared resources whereas 
%% producer_consumer does NOT.  
%% Consider redacting producer-consumer if space is limited.

Figure \ref{Figure:producerconsumer} illustrates the benefits of CR on the
\texttt{producer\_consumer} benchmark from the \texttt{COZ} package \cite{sosp15-curtsinger}.  
The benchmark implements a bounded blocking queue by means of a
pthread mutex, a pair of pthread condition variables to signal \textit{not-empty} 
and \textit{not-full} conditions, and a standard C++ \texttt{std::queue<int>} container
for the values. (This implementation idiom -- a lock; a simple queue; and two condition variables -- 
is common).  Threads take on fixed roles, acting as either producers or consumers. 
The benchmark spawns $N$ concurrent threads, each of which loops, producing or
consuming according to its role.  We fix the number of consumers at 3 threads and vary the number
of producers on the X-axis, modeling an environment with 3 server threads and a variable number
of clients.  We report the number of messages conveyed at the end
of a 10 second measurement interval, taking the median of 7 distinct trials. 
The queue bound was 10000 elements.

Under a classic FIFO lock,
when the arrival rate of producers exceeds that of consumer threads, 
producers will acquire the lock and then typically find the queue is full and thus block 
on the condition variable, releasing the lock. Eventually they reacquire the lock, 
insert the value into the queue, and finally release the lock \footnote{
The condition variable implementation used in these experiments provides FIFO order.}.
Each conveyed message requires 3 lock acquisitions -- 2 by the producer and one by the consumer.
The critical section length for producers is artificially increased by futile acquisitions
where the producer immediately surrenders the lock and blocks on the condition variable.  
When the condition variable is subsequently signaled, the producer moves to the tail of the lock queue.
Producers typically block 3 times : first on arrival to acquire the lock; on the condition variable; and
on reacquisition of the lock.  Ownership of the lock circulates over all participating threads.  
The queue tends to remain full or nearly so, and consumers do not need to wait on the 
\emph{not-empty} condition variable.  

Under a CR lock we find the system tends to enter a desirable ``fast flow'' mode 
where the futile acquisition by producers is avoided and each conveyed message requires only 
2 lock acquisitions.  Threads tend to wait on the mutex instead of on condition variables.  
Given sufficient threads, ownership continuously  circulates over a small stable balanced set of 
producers and consumers.  (As usual, long-term fairness enforcement ensures eventual participation
of all threads).  We note that CR's mode of benefit for the other benchmarks
involves competition for fixed shared resources, whereas producer\_consumer demonstrates
benefits from reduced lock acquisition rates and hold times
%% OPTIONAL  
\footnote{Medhat et al. \cite{ipdps14-medhat} explored the interaction of waiting policies
on CPU sleep states for producer-consumer applications.}. 

%% Waiting is confined to mutex instead of condvars

\Invisible{Tends to converge to; stable state; stable mode; tend; trend; evolve toward;
devolve; degenerate to; homeostatis;} 

\subsection{keymap benchmark} 

The \texttt{keymap} benchmark in Figure \ref{Figure:keymap} spawns set of concurrent threads, each of which 
loops executing a critical section followed by a non-critical section.  At the end of a 10-second
measurement interval the benchmark reports the aggregate throughput as the total number of loop
iterations completed by all the threads.   The non-critical section advances a C++  
\texttt{std::mt19937} pseudo-random number generator 1000 times.  The critical section 
acquires a central lock and then picks a random index into its thread-local \textit{keyset}
array.  Each keyset array contains 1000 elements and is initialized to random keys 
before the measurement interval.  With probability $P=.9$ the thread then extracts
a key from its keyset and updates a central C++ \texttt{std::unordered\_map<int,int>} instance
with that key.  Otherwise the thread generates a new random key in the range
$[0,10000000)$, updates the keyset index with that key, and then updates the shared map.  
All pseudo-random generators are thread-local and uniform.
%% OPTIONAL ...
To reduce allocation and deallocation during
the measurement interval, we initialize all $10000000$ keys in the map prior to spawning
the threads. 

\texttt{Keymap} models server threads with short-lived session connections and moderate
temporal key reuse and memory locality between critical sections executed by a given thread. 
There is little or no inter-thread CS access locality or similarity, however.  
Threads tend to access different regions of the CS data.  
The NCS accesses just a small amount of memory, and CR provides benefit by moderating 
inter-thread competition for occupancy of CS data in the shared LLC.   

\subsection{LRUCache} 
Figure \ref{Figure:LRUCache} shows the results of the \texttt{LRUCache} benchmark.
\texttt{LRUCache} is derived from the \texttt{keymap} benchmark, but instead of
accessing a shared array, it executes lookups on a shared LRU cache.  
We used the \texttt{SimpleLRU} C++ class from the \texttt{CEPH} 
distributed file system, found at  \url{https://github.com/ceph/ceph/blob/master/src/common/simple_cache.hpp}.
SimpleLRU uses a C++ \texttt{std::map} -- implemented via a red-black tree -- protected by a single
mutex.  The map associates 32-bit integer keys with a 32-bit integer cached value.  
Recently accessed elements are moved to the 
front of doubly linked list, and excess elements are trimmed from the tail of that list in order
to enforce a limit on the number of elements.   On a cache miss we simply install the key itself
 as the value associated with the key; miss overheads are restricted to removing and inserting 
elements into the \texttt{std::map}.  
We set the maximum capacity of the SimpleLRU cache at 10000 elements.  
The key range was $[0,1000000)$.  Like \texttt{keymap}, the key set size was 1000 elements.  
The key set replacement probability was $P=0.01$.  
Whereas \texttt{keymap} demonstrated inter-thread competition for occupancy of the array 
shared hardware LLC, threads in \texttt{LRUCache}, compete for occupancy in the software-based 
LRU cache.  Concurrency restriction reduces the miss rate and destructive interference in the software
LRU cache.  The LRU cache is conceptually equivalent to a small shared hardware cache
having perfect (ideal) associativity  
\footnote{In \texttt{LRUCache} it is trivial to collect displacement statistics
and discern self-displacement of cache elements versus displacement caused by other 
threads, which reflects destructive interference.}. 

\subsection{perl benchmark}
In Figure \ref{Figure:perl} we report the performance of the \texttt{RandArray} 
benchmark, ported to the perl language and using arrays with 50000 elements.  
We used an unmodified version of perl 5.22.0, which is the default version provided
by the Solaris distribution.  
Perl's lock construct consists of a pthread mutex, a pthread condition variable, 
and an owner field.  Threads waiting on a perl lock will wait
on the condition variable instead of the mutex, and the underlying mutex rarely encounters
contention, even if the lock construct is itself contended.  CR on the mutex
would provide no benefit for such a design.  Instead, we apply CR via the condition variable
by way of LD\_PRELOAD interposition on the pthread interface.  
The experiment uses a classic MCS lock for the mutex, providing FIFO order.
We modified the pthread condition variable construct to allow both FIFO ordering (unless
otherwise stated, all condition variables used in this paper provide strict FIFO
ordering) and a \emph{mostly-LIFO} queue discipline, which provides CR.   
Each condition variable has a list of waiting threads.  The default FIFO mode
enqueues at the tail and dequeue from the head.  In mostly-LIFO mode we use a 
biased Bernoulli trial to determine if a thread is to be added at the head or tail 
by the \texttt{wait} operator.  With probability $999/1000$ we prepend to the head,
and 1 out 1000 \texttt{wait} operations will append at the tail, 
providing eventual long-term fairness.  For simplicity, we used a waiting strategy with
unbounded spinning.  We see that the mostly-LIFO mode provides better performance 
at about 5 threads, due to reduced LLC pressure.  Performance for both condition 
variable policies fades at 128 threads where we have more threads than logical processors.  
Since perl is interpreted, the absolute throughput rates are far below that of 
\texttt{RandArray}.  We also tried to recapitulate \texttt{RandArray} in the python 
language, but found that scalability was limited by Global Interpreter Lock (GIL).  

\subsection{Buffer Pool benchmark} 
The Buffer Pool benchmark use a central shared blocking buffer pool implemented with a pthread mutex, 
a \emph{NotEmpty} pthread condition variable, and a C++ \texttt{std::deque}
which contains pointers to available buffers.  \Invisible{The pool uses a LIFO allocation policy.} 
All buffers are 1MB in length and the pool is initially provisioned with 5 buffers. 
The pool allows unbounded capacity but for our benchmark will never contain more than 5 
buffers at any given time.  
The benchmark spawns $T$ concurrent threads, each of which loops as follows:
allocate a buffer from the pool (possibly waiting until buffers became available); 
exchange 500 random locations from that buffer with a private thread-local buffer; 
return the buffer to the pool; update 5000 randomly chosen locations in its private buffer.  
At the end of a 10 second measurement interval the benchmark reports the aggregate 
iteration rate (loops completed) per second.  We report the median of 5 runs.  
In Figure \ref{Figure:BP} we vary the number of threads on the X-axis and report 
throughput on the Y-axis.  The pthread mutex construct was implemented as a classic MCS lock and
the condition variable implementation used an explicit list of waiting threads, and
allowed us to vary, via environment variable, the probability $P$ which controls whether
a thread would be added to the head or tail of the list.  Both the mutex and
condition variable employed an unbounded spinning waiting policy.  Our sensitivity analysis shows
the performance at a number of $P$ values.  If the probability to append is 1
then we have a classic FIFO policy and if the probability is 0 we have a LIFO policy,
otherwise we have a mixed append-prepend policy.  As shown in the Figure, pure 
prepend (LIFO) yields the best throughput.  As we increase the odds of appending, 
throughput drops.  A mostly-prepend policy (say, 1/1000) yields most of the 
throughput advantage of pure LIFO, but preserves long-term fairness.  
CR results in fewer circulating threads in a given period, which in turn means
fewer buffers being accessed, lower LLC pressure and miss rates, and higher 
throughput rates. 

We also experimented with a buffer pool variant using a POSIX pthread semaphore 
instead of a condition variable, where threads waiting for a buffer to become 
available will block on the semaphore instead of on the condition variable.   
We then implemented an LD\_PRELOAD interposition module that intercepts 
\texttt{sem\_wait} and \texttt{sem\_post} semaphore operations.  
Our semaphore implementation uses an explicit list of waiting threads and was 
equiped to allow the append-prepend probability $P$ to be controlled via 
environment variable.  The results were effectively identical 
to those shown in Figure-\ref{Figure:BP}, showing that CR provided via semaphores is effective.

%% footnote or paragraph ?
%% Related: FOLLY LifoSem -- http://queue.acm.org/downloads/applicative/bmaurer.pdf

We note that the FaceBook \texttt{FOLLY} library --
\url{https://github.com/facebook/folly/blob/master/folly/LifoSem.h} --
includes the \texttt{LifoSem} LIFO Semaphore construct.  They claim: 
\textit{LifoSem is a semaphore that wakes its waiters in a manner intended to
maximize performance rather than fairness.  It should be preferred
to a mutex+condvar or POSIX sem\_t solution when all of the waiters
are equivalent}.  LifoSem uses an always-prepend policy for strict LIFO admission,
whereas our approach allows mixed append-prepend ensuring long-term fairness,
while still providing most of the performance benefits of LIFO admission.
By providing long-term fairness, semaphores augmented with CR are acceptable
for general use instead of limited specific circumstances, 
as would be the case for LifoSem. 

CR can also provide related benefits for thread pools, where idle threads awaiting
work block on a central pthread condition variable.  With a FIFO fair condition variable,
work requests are dispatched to the worker threads in a round-robin fashion.
Execution circulates over the entire set of worker threads.  
With a mostly-LIFO condition variable, however, just the set of worker threads
needed to support the incoming requests will be activated, and the others can
remain idle over longer periods, reducing context switching overheads.  
\Invisible{Observed benefit of CR in thread pools with \texttt{libuv}, 
which underlies \texttt{node.js}.} 

\Invisible{Effectively equivalent; effectively identical.}

%% ===================================================================
%% Retain for reference ...
%% Consider : sidewaysfigure
%% See also : LaTex tricks from SPAA 2014 BA -- CacheResidencyUnfairness

%% \begin{figure*}[hbtp]
%% \begin{center}
%% %% \epsfxsize=2.20in \epsfbox{misses.eps}
%% %% \resizebox{....}
%% \includegraphics[angle=270,origin=c,width=16cm]{plot-ringer/plot.eps}
%% \end{center}
%% \caption{Single-threaded ring traversal rates} 
%% \label{figure:ringer}
%% \end{figure*}
%% 
%% \begin{figure*}[hbtp]
%% \begin{center}
%% %% \epsfxsize=2.20in \epsfbox{misses.eps}
%% %% \resizebox{....}
%% \includegraphics[angle=270,origin=c,width=16cm]{plot-avl/plot.eps}
%% \end{center}
%% \caption{AVL tree throughput with 4 threads} 
%% \label{figure:avl} 
%% \end{figure*}
%% ===================================================================

%% \input{discussion} 

\section{Discussion} 

MCSCR is robust under varying load and adapts the size of the 
ACS quickly and automatically, providing predictable performance.  
\Invisible{
For instance if the system load increases and we find more ready threads
than logical processors, then the system will start to multiplex the logical
processors via involuntary preemption.  Contention on an already contended
lock may increase because of lock holder preemption.  The MCS chain will
grow, but MCSCR responds immediately by increased culling, reducing
the size of the ACS to a level appropriate for the new system load. } 
The implementation of MCSCR is entirely in user-space and requires no
special operating system support.  
No stateful adaptive mechanisms are employed, resulting
in more predictable behavior and faster response to changing conditions. 
The only tunable parameter, other than the spin duration, is how frequently
the unlock operator should pick the eldest thread from the passive set, which
controls the fairness-throughput trade-off.  

\Invisible{Performance does not degrade as load increases, mitigating
the need for load-specific and application-specific tuning.} 
\Invisible{Parameter Parsimony} 

%% OPTIONAL ...
Involuntary preemption, which typically manifests when there are more ready threads
than logical CPUs and the operating system is forced to multiplex the 
CPUs via time slicing, can cause both lock holder preemption and lock waiter preemption.
The former concern can be addressed in various ways \cite{transact16-dice}, including deferred
preemption via schedctl.  Lock holder preemption can also be mitigated by a ``directed yield'' 
facility, which allows a running thread to donate its time slice to a specified thread 
that is ready but preempted.  This allows threads waiting on a lock to grant CPU time 
to the lock holder, reducing queueing and convoying.  Lock waiter preemption entails handoff 
by direct succession to a thread that is waiting but was preempted.  MCS, for instance,
is vulnerable to lock waiter preemption \cite{blog-preemptiontolerantmcs} --
we say MCS is \emph{preemption intolerant} -- whereas simple TAS locks are not.

\Invisible{
* Directed Yield = YieldTo or SwitchTo; 
* Orbit; trajectory; path; cycle; circuit; circumference; 
* Epicycle
} 

%% OPTIONAL ...
%% Possible footnote 
It is sometimes possible to use schedctl with direct succession locks to avoid handoff to a 
preempted waiter.  The thread in unlock() examines the schedctl state of the tentative
successor.  If that thread was preempted, then it picks some other thread instead.  
Early experiments suggest that it is helpful to use the schedctl facility to detect 
preempted threads on the MCS chain.  The unlock() operator can check for and evict 
such threads from the chain, forcing them to recirculate and ``re-arrive'' after 
they are eventually dispatched, making schedctl augmented MCS far more tolerant
of waiter preemption and reducing the incidence of ownership being transferred
to a preempted thread.  

%% No machine learning !
%% TAG:SLEEP
%% De-emphasize sleep-idle state mode of benefit
%% CR also acts to reduce voluntary context switch rates.  
CR also actively reduces the voluntary context switch rate. 
Since the passive set can remain stable for prolonged periods, threads in the passive set
perform less voluntary context switching (park-unpark activity), 
which in turn means that the CPUs on which those threads 
were running may be eligible to use deeper sleep states and enjoy reduced power 
consumption and more thermal headroom for turbo mode  
%% OPTIONAL ...
\footnote{Similarly, effective swapping requires that the balance set not change too rapidly.}. 
Relatedly, CR acts to reduce the number of threads concurrently spinning on a given 
lock, reducing wastage of CPU cycles.  Voluntary blocking reduces the involuntary
preemption rate as having fewer ready threads results in less preemption. 
That is, concurrency restriction techniques may reduce involuntary preemption rates by
reducing the number of ready threads competing for available CPUs.  This
also serves to reduce lock-holder preemption and convoying 
\footnote{Solaris provides the \emph{schedctl}\cite{schedctl,blog-invertedschedctl} facility to request
advisory deferral of preemption for lock holders -- lock-holder preemption avoidance. 
Edler \cite{edler} proposed a similar mechanism.  Schedctl can also be used
to detect if the lock holder itself is running, allowing better informed 
waiting decisions. We did not utilize schedctl in the experiments reported in
this paper.}.

\Invisible{
CR actively reduces park-unpark voluntary context switching rates
by keeping the ACS stable and minimal, in turn .  Reducing the park-unpark rate
also acts to reduce the CPU transitions between idle and running.} 

%% OPTIONAL ...
%% TAG:SLEEP
\Invisible{
Absent CR, lock ownership can circulate over a larger number of threads (CPUs) 
in a given period.   Some of those threads may wait by blocking in the kernel, 
potentially making their CPUs become idle.  Rapid circulation of the lock over 
this larger circulating set may cause CPUs to shift between idle and non-idle more 
rapidly, both incurring latency in the idle to non-idle transition, and also 
prohibiting the CPUs underlying the ACS from reaching deeper energy-saving state.  
Those deeper sleep states also enable more aggressive turbo mode \cite{TurboDiaries} 
for other sibling cores, allowing threads on those cores to run faster.

%% OPTIONAL ...
%% TAG:SLEEP
By minimizing the size of the ACS, we tend to fully utilize the set of CPUs
hosting the ACS threads.  These CPUs do not become idle, and thus do not 
incur latency exiting deeper sleep states.  Furthermore other CPUs not hosting ACS 
threads may enjoy longer idle periods and deeper idle sleep states, thus improving 
energy consumption and possibly allowing more available thermal headroom for either
other unrelated threads, or to allow members of the ACS to run at higher clock
rates. 
}

\Invisible{Idle state; sleep state; park CPU; C-state; P-state;} 

%% OPTIONAL ...
%% Possible footnote in conclusion section 
A common admonition is to never run with more threads than cores.  
This advice certainly avoids some types of scaling collapse related to core-level resource competition,
but is not generally valid, ignoring the potential benefit of memory-level parallelism (MLP),
threads that alternate between computation and blocking IO, etc.  
Many applications achieve peak throughput with far more threads than cores.  
Such advice also assumes a simplistic load with just one application, whereas 
servers may run in conditions of varying load and multi-tenancy, with multiple 
concurrent unrelated and mutually-unaware applications.
Even within a single complex application we can find independent components 
with their own sets of threads, or thread pools.  CR provides particular benefit 
in such real-world circumstances. 
\Invisible{DevOps} 
\Invisible{Bursty; Transient; offered load; steady-state load; fixed load;}

%% OPTIONAL ...
\Invisible{
We observe that in some situations, no speedup is to be had even as we increase 
the number of threads, even though the NCS length might be far larger than the 
CS length.  This happens frequently when the CS and NCS lengths
are short, in which case lock administrative overheads may dominate throughput. 
In practice, locks are never ideal.  Speedup can be impaired by coherence traffic
-- sharing that involves writes, resulting in invalidation and misses -- because of 
data accesses in the critical section.  Lock metadata access  will also induce 
coherence traffic.  
Other overheads related to handover latency include the branch 
mispredict stalls that are incurred when a thread exits an busy-wait loop. 
\Exclude{It is not uncommon to observe a slight transient ``dip'' in performance at low 
thread counts, when the benefit increased parallelism has not overcome coherence overheads.  
This behavior can be seen in Figure \ref{Figure:dtlb} at 2 threads with MCS.} 
}

\section{Related Work} 

%% Related work section should appear either immediately before conclusion
%% section or immediately after the introduction.
%% Most closely resembles that of ...

%% OPTIONAL: cull following citation list ...
Locks continue to underpin most applications and remain a key synchronization construct.
They remain the topic of numerous recent papers 
\cite{Eastep-smartlocks,locks-ols,cacm15-bueso,ieeetocs15-cui,ppopp15-chabbi,
apsys15-Kashyap,ppopp16-chabbi,ppopp16-wang,transact16-dice,usenixatc16-guiroux,sac16-gustedt,
usenixatc16-falsafi,ppopp16-ramalhete}. 

Our work is most closely related to that of Johnson et al. \cite{asplos10-johnson-decoupling},
which also addresses performance issues arising from overthreading, using load 
and admission control to bound the number of threads allowed to spin concurrently 
on contended locks.  Their key contribution is controlling the spin/block waiting decision 
based on load.  If the system is overloaded, in which case there are more ready 
threads than logical CPUs, then some of the excess threads spinning on locks are 
prompted to block, reducing futile spinning and involuntary preemption.  
Their scheme operates only when the number of ready threads exceeds the number
of logical CPUs, and some of those threads are spinning, waiting on locks,
whereas ours responds earlier, at the onset of contention, and controls the
size of the each individual lock's active circulating set.   
This allows our approach to moderate competition for other shared resources 
such as residency in shared caches and pipelines.
Their approach operates system-wide and requires a daemon thread to detect and 
respond to contention and load whereas ours uses timely decentralized local 
per-lock decisions and is easier to retrofit into existing lock implementations.  
They also requires locks which are abortable, such as TP-MCS \cite{hipc05-he}.  
Threads that abort -- shift from spinning to blocking -- must ``re-arrive'', with 
undefined fairness properties.  
Their approach can easily leave too many spinning threads with ensuing intra-core
competition for pipelines, whereas ours is more appropriate for modern multicore processors.  
We treat the spin/block waiting policy as a distinct albeit important concern.  
The two ideas are orthogonal, and could be combined.

%% Regarding novelty : We will amend the related work section                                               
%% to more clearly differentiate our approach.
\Invisible{
* Oversubscription; addresses overthreading; 
* Deleterious; unprofitable; futile; 
* avoid spinning when there are too many ready threads;
* reduces preemption and futile spinning; 
* spin/block decision based on the number of ready threads vs number of CPUs;
* Load is the number of ready threads or numberready / numbercpus = utilization;
* load vs utilization; offered load
* Alternative definition : load = availability of ready CPUs
* Their approach only active-operative when the number of ready threads exceeds 
  the number of logical CPUs.
* Their approach presumes that all locks in the system participate in and adhere 
  to their protocol.
* addressing contention and scheduler-level competition 
  for those logical CPUs by ready threads, and reducing inopportune preemption.   
* They avoid futile and deleterious spinning when there are more ready 
  threads than CPUs by prompting spinning threads to switch to blocking.  
* Their approach is driven by load, and controls the spin/block waiting policy.
* excess spinning threads are prompted to block.  
* Ours responds directly to contention and controls the size of the active circulating set.   

Comparison -- Johnson:
* system-wide; centralized
* requires abortable lock
* wait strategy only : block vs spin
* shift excess waiting spinners to blocking
* driven by load/cpus; controls spin vs block
Ours 
* per-lock local decentralized decisions
* drive by contention; controls ACS size

Culled sentences:
They also require the ability to detect overload whereas ours does not.
To maximize effectiveness, all synchronization constructs in the system that might 
cause threads to wait should adhere to their protocol.   

} 

Chadha et al. \cite{cases12-chadha} identified cache-level thrashing as a 
scalability impediment and proposed system-wide concurrency throttling.  
Throttling concurrency to improve throughput was also suggested by Raman et al. \cite{pldi11-raman} 
and Pusukuri et al. \cite{iiswc11-pusukuri}.   
Chandra et al. \cite{hpca05-chandra} and Brett et al. \cite{ipdpsw13-brett}  analyzed the 
impact of inter-thread cache contention.
Heirman et al. \cite{hpca14-heirman} suggested intentional undersubscription of threads
as a response to competition for shared caches. 
Mars et al. \cite{cgo10-mars} proposed a runtime environment to reduce cross-core interference. 
Porterfield et al. \cite{ipdpsw13-porterfield} suggested throttling concurrency 
in order to constrain energy use.  
Zhuravlev et al. \cite{csurv12-zhuravlev} studied the impact of kernel-level scheduling
decisions -- deciding which and where to dispatch ready threads -- on shared resources,
but did investigate the decisions made by lock subsystems.  
Cui et al. \cite{lsap11-cui} studied lock thrashing avoidance techniques in the 
linux kernel where simple ticket locks with global spinning caused scalability
collapse.  They investigated using spin-then-park waiting and local spinning, but
did not explore CR. 

Like our approach, \emph{Cohort locks} \cite{topc15-dice} explored the trade-off 
between throughput and short-term fairness.  Cohort locks restrict the active 
circulating set to a preferred NUMA node over the short term.  They sacrifice
short-term fairness for aggregate throughput, but still enforce long-term fairness.
NUMA-aware locks exploit the inter-socket topology, while our approach focuses 
on intra-socket resources.  
%% OPTIONAL ...
The NUMA-aware HCLH lock \cite{HCLH} edits the nodes of a queue-based lock in a fashion 
similar to that of MCSCR, but does not provide CR and was subsequently discovered 
to have an algorithmic flaw.

Johnson et al. \cite{damon09-johnson} and Lim et al. \cite{tocs93-lim} explored the trade-offs between 
spinning and blocking.  

Ebrahimi et al. \cite{micro11-Ebrahimi} proposed changes to the system scheduler,
informed in part by lock contention and mutual inter-thread DRAM interference,
to shuffle thread priorities in order to improve overall throughput.   

%% OPTIONAL ...
Hardware and software transactional memory systems use \emph{contention managers} to 
throttle concurrency in order to optimize throughput \cite{spaa08-yoo}.  The issue
is particularly acute for transactional memory as failed optimistic transactions
are wasteful of resources.  

%% OPTIONAL ...
Various hardware schemes have been proposed to mitigate LLC thrashing,
but none are available in commonly available processors \cite{js04-suh}.  
Intel \cite{Intel-CAT} allows static partitioning of the LLC in certain
models designed for real-time environments.  

\section{Conclusion}

Modern multicore systems present the illusion of having a large number
of individual independent ``classic'' processors, connected via shared memory.  This abstraction, 
which underlies the symmetric multiprocessing SMP programming model, 
is a useful simplification for programmers.   
In practice, however, the logical processors comprising these multicore systems
share considerable infrastructure and resources.  Contention for those shared 
resources manifests in surprising performance issues.  

\Invisible{
*  Destructive interference means we often face negative-sum situations. 
*  sub-additive
*  Illusional; notional; illusory; facade; 
*  Locks are in the business of medium-term scheduling. 
*  Performance isolation failure;
*  cite spaa14-dice
} 

\Invisible{
Multicore systems are fundamentally a deceit.  Most of the time we live happily 
with the ``SMP'' illusion that we have a large number of independent processors. 
MLP and lowered communication costs active as palliative factors. 
But sometimes we have to face reality and deal with the fact 
that there are really lots of shared resources, subject to contention and 
even destructive interference, under the facade. 
} 

We describe a lock admission policy -- concurrency restriction -- that is
intentionally unfair over the short term.  Our algorithm intentionally
culls excess threads -- supernumerary threads not required to sustain contention --
into an explicit passive set.  CR moderates and reduces the
size of the active circulating set, often improving throughput relative to 
fair FIFO locks.  
Periodically, we reschedule, shifting threads between the active and passive
sets, affording long-term fairness.  CR conserves shared resources and
can reduce thrashing effects and performance drop that can occur when
too many threads compete for those resources, demonstrating that
judiciously managed and intentionally imposed short term unfairness can
improve throughput.  
CR provides a number of modes of benefit for the various types of 
shared and contended resources. 
We further show the subtle interplay of waiting policy, which must be
carefully selected to fully leverage CR.  

While scalability collapse is not uncommon, it remains a challenge
to characterize which shared resources underly a drop in performance.
The analysis is difficult and in our experience, multiple resources are
often involved \footnote{Suggesting the need for enhanced hardware performance facilities
to detect excessive competition for shared resources.}. 
While CR typically does no harm, it is also difficult to determine in 
advance if CR will provide any benefit.
\Invisible{However, since CR typically does no harm, the decision to use it is simple.}   
%% OPTIONAL ...
CR gates access to the resources
involved in scalability collapse by moderating access to locks -- an unrelated resource.
In the future we hope to employ more direct means to measure and control scalability 
collapse.  
Locks remain convenient, however, and detecting oversubscription (contention)
is relatively simple compared to determining when some of the complex hardware resources
are oversubscribed. 
Contention is a convenient but imprecise proxy for overthreading.  
%% Contention is; contention serves as; ...

\Invisible{
*  graceful; predictable; robust; automatic; autonomic;
*  adjusts automatically and promptly to varying load;  
*  respond; response; react; adapt; adjust; lag; latency; promptly;
   reaction time; response time; 
*  Unsatisfying;
*  CR works in concert with both operating system and hardware 
*  under high load -- handles preemption gracefully; 
   under varying load -- fast response time
}

%% \subsection{Contributions}
%% Enumerate contributions of the paper ...

\subsection{Future Work} 

Throttling in current CR designs is driven by the detection of contention.
In the future we hope to vary the admission rate (and the ACS size) in order 
to maximize lock transit rates, possibly allowing non-working conserving 
admission \cite{SIF}.  This attempts to 
close the performance gap between \emph{saturation} and \emph{peak} shown in 
Figure \ref{Figure:CR-Graph}. 
%% plan to apply; plan to further explore
\Invisible{ We also plan to further explore the application of intentionally 
unfair CR-based activation policies to semaphores and the \texttt{pthread\_cond} 
condition variable construct, tending to wake the most recently arrived threads.  
This approach shows promise for pools of worker threads where
idle threads wait on a central condition variable.}
We also intend to explore energy-efficient locking in more detail, 
and the performance advantages of CR on energy-capped systems.  

Classic CR is concerned with the size of the ACS.  But we can easily
extend CR to be NUMA-aware by taking the demographics of the ACS 
into account in the culling criteria. For NUMA environments we prefer the ACS to 
be homogeneous and composed of threads from just one NUMA node.  This reduces
the NUMA-diversity of the ACS, reduces lock migrations and improves performance.  
Our \textbf{MCSCRN} design starts with MCSCR, but
we add two new fields: the identity of the currently preferred ``home'' NUMA
node, and a list of remote threads.  At unlock-time, the owner thread inspects 
the next threads in the MCS chain and culls remote threads from the main chain 
to the remote list.  A thread is considered remote if it runs on some node 
other than the currently preferred node.  Periodically, the unlock operator also
selects a new home node from the threads on the remote list, and drains threads
from that node into the main MCS chain, conferring long-term fairness.  
If we encounter a deficit on the main list at unlock-time, then we simply 
reprovision from the remote list.   

\Invisible{
*  cull; splice; extract; excise; demote; remove; move; shift; sift; redact; resect 
*  relegate; demote; 
*  ACS = enabled set; PS = disabled set;  
*  Surplus; excess; supernumerary; nimiety; overabundance; superabundance; 
   surfeit; superfluity; redundant threads; 
*  reduce NUMA-diversity of ACS; homogenize
} 

Early experiments with NUMA-aware CR show that
MCSCRN performs as well as or better than CPTLTKTD\cite{topc15-dice}, the best known cohort lock.  
In addition, cohort locks require one node-level lock for each NUMA node.  Because
of padding and alignment concerns to avoid false sharing, those node-level locks themselves
are large.  
\Invisible{MCSCRN avoids that -- the lock size is fixed and small.}
%% MCSCRN locks are small and of fixed size.
Unlike cohort locks, MCSCRN locks are small and of fixed size
In the uncontended case, cohort locks require acquisition of both the node-level and top-level,
although a fast-path can be implemented that tries to avoid that overhead by
opportunistically bypassing the node-level locks under conditions of no or light contention
when cohort formation is not feasible. 
MCSCRN is non-hierarchical, and avoids that concern, always using the fast-path. 
The system tends to converge quickly to a steady-state where
the arriving threads are largely from the home node, so accesses to lock metadata
elements avoids inter-node coherence traffic.  
Finally, we note that it is a challenge to implement polite spin-then-park
waiting in CPTLTKTD, but it is trivial to do so in MCSCRN.  
MCSCRN will be the topic of a subsequent publication. 

Concurrency restriction has also be used to reduce virtual memory pressure and
paging intensity. Just as the LLC is a cache backed by DRAM, DRAM is a cache
backed by the system's paging resources.  
On SPARC, we have also found CR can reduce pressure and destructive interfence
with a process's translation storage buffer (TSB).  We hope to include the results 
of these experiments in future publications.  

\acks

We thank Alex Kogan, Doug Lea, Jon Howell and Paula J. Bishop for useful discussions.

%=========================================================================
%  Bibliography
%=========================================================================

%% When using Bibtex, the following form may be used.
%% ----------------------------------------------
%% consider natbib
%% \bibliographystyle{plain}
%% \bibliographystyle{abbrv}
%% \bibliographystyle{abbrvnat}
%% \bibliography{bib.bib}
%% \bibliography{bib}
%% \newcommand{\bibfont}{\footnotesize} 
%% \newcommand{\bibfont}{\9pt}         % Bibliography's font size;
%% \setlength{\bibhang}{4ex}                    % Indent of Bibliography entries;
%% \setlength{\bibsep}{3pt}                     % Separation between BiB entries;
%% \renewcommand*{\bibfont}{\footnotesize}
%% ----------------------------------------------
%% \bibliographystyle{abbrv}
%% {\footnotesize \bibliography{bib}} 
%% \small
%% ----------------------------------------------
%% \bibliographystyle{abbrv}
%% \bibliographystyle{abbrvnat} 
%% ----------------------------------------------
%% \newcommand{\bibfont}{\9pt}         % Bibliography's font size;
%% ----------------------------------------------

\renewcommand{\bibfont}{\scriptsize} 
\bibliographystyle{abbrvnat}            %% abbrv vs abbrvnat
{\scriptsize \bibliography{main}}

%=========================================================================
%  Appendix
%=========================================================================

%% \pagebreak \small
%% \small
%% \appendix vs \begin{appendix} and \end{appendix} 

\appendix
%% Use ambient normal size 
%% \footnotesize
%% \scriptsize
%% \footnotesize

\section{Additional lock formulations that provide concurrency restriction} 

%% To illustrate the ease with which locks providing CR can be constructed, we 
%% provide additional examples.  
We provide additional examples to illustrate generality and show that other locks providing
concurrency restriction can be constructed. 

%% \Invisible{
%% Construction of CR variants; 
%% Construction of other locks with CR;
%% Variations that provide CR; 
%% Other lock formulations that provide CR; 
%% Additional lock formations that provide CR;
%% Additional lock constructions that provide CR;
%% Algorithmic sketches
%% }
 
\subsection{LOITER Locks} 

%% Possibly move following para to "waiting policy" section ...
Simple TAS or more polite test-and-test-and-set spin locks can be deeply unfair.  
A thread can repeatedly barge in front of and bypass threads that have
waited longer.  A simple TAS lock without back-off can also suffer 
from considerable futile coherence traffic when the owner releases the lock and 
the waiting threads observe the transition and $N$ such spinning threads \emph{pounce}, 
trying to obtain ownership via an atomic instruction, producing a
\emph{thundering herd} effect.  $N-1$ will fail, but in doing 
so force coherence traffic on the underlying cache line.
As such, modern TAS locks 
are typically augmented with randomized back-off, which reduces coherence traffic 
from polling and also reduces the odds of futile attempts to acquire the lock.  
Back-off strives to balance those benefits against reduced lock responsiveness.  
Longer back-off periods entail longer possible ``dead time'' where the lock has
been released but the waiting threads have not detected that transition 
\footnote{Arguably, back-off is not work conserving.}.  
Traditional randomized back-off for TAS locks is \emph{anti-FIFO} in the sense that threads that 
have waited longer are less likely to acquire the lock in unit time.  
Absent remediation, such back-off may partition threads into those that
wait for long periods and those that wait for short periods and circulate rapidly
\footnote{The back-off can also provide inadvertent and unintended but beneficial
concurrency restriction.}. 

\Invisible{
*  Unfair : admission order is decoupled from arrival order
*  Barge; Overtake; bypass; jump 
*  Lunge; Pounce;
*  Dead time; lag; latency; hand-over responsiveness; conveyance; 
*  Admission order is decoupled from arrival order
}

\Invisible{List of LOITER optimizations: inverted schedctl for spinners; 
Limit number of concurrent spinners; wakeup-unpark deferral and avoidance; 
fence elision; anti-spinning by standby thread; fairness enforcement;} 

%% OPTIONAL : consider as footnote ...
%% Partial remedy; mitigation; relief; 
Anderson \cite{anderson-spinning} suggested the following partial remedy for 
the thundering herd effect.  After a spinning thread observes the lock 
is free, it delays for a short randomized period and then re-checks the lock.  
If the lock is not free then the thread resumes normal spinning.  Otherwise the 
thread attempts the atomic operation to acquire the lock. 
%% suggested a randomized delay and re-check of the lock after it was observed free 
%% but before the atomic operation in order to reduce this \emph{thundering herd} effect.
This technique has lost favor in modern TAS locks, but is useful when
used in conjunction with MWAIT.

Fairness of TAS locks is further determined by platform-specific aspects of the system
such as the underlying hardware arbitration mechanism for cache lines.  
On some platforms, threads running ``near'' the most recent owner -- near in 
the system topology -- may enjoy a persistent statistical advantage acquiring the lock,
dominating ownership.  
On some platforms, threads on the home node of the memory 
underlying the lock will have a persistent advantage.  Somewhat perversely, such 
behavior can be NUMA-friendly over the short-term as it tends to reduce lock migrations.  
The unfairness can persist for long periods, however.  

\Invisible{
*  Lock Philopatry; propinquity; inertia
*  Matthew effect; virtuous circle; positive feedback; self reinforcing;
   The rich get richer; amplification; 
   The faster a thread runs, the better relative residency, and the
   faster it will run in the future. 
*  Confounding factor :
   Brief announcement: persistent unfairness arising from cache residency imbalance
   SPAA 2014; Dice
*  Fairness measures deviation of admission order from ideal strict FIFO.
} 
 
\Invisible{On Intel and SPARC processors if a thread $T$ releases a TAS
lock, siblings of $T$ from the same core or NUMA node enjoy a statistical advantage 
in next acquiring the lock.  Threads from a given node may dominate ownership of a 
lock for extended periods. } 

Despite these disadvantages, TAS locks confer a key benefit:
the lock is never passed to a preempted thread as might be the case with MCS  
%% OPTIONAL ...
%% \footnote{Kontothanassis et al. \cite{tocs97-kontothanassis} proposed variants 
%% of MCS that avoid this issue.}.  
This reduces undesirable convoying behavior and latencies waiting for
a ready but descheduled thread to again be dispatched onto a CPU.  
Furthermore, waiting threads do not need to register or otherwise make
themselves visible to threads performing the unlock operation, reducing
administrative overheads and coherence costs related to lock metadata.  
As such, these locks perform better under mixed load, and in particular when the
number of runnable threads exceeds the number of logical CPUs.  They also
have very low latency hand-off under light or no contention.  

We design a simple TAS lock enhanced with CR as follows.  Our new 
\textbf{LOITER} (\underline{L}ocking : \underline{O}uter-\underline{I}nner 
with \underline{T}h\underline{R}ottling)
lock has an \emph{outer} TAS lock.  Arriving threads try to obtain the outer lock using
a bounded spin phase -- busy waiting -- with randomized back-off.  If they 
acquire the outer lock, they can enter
the critical section.  We refer to this as the fast-path.  If the spinning attempt 
fails, control then reverts to an \emph{inner lock}.  An MCS lock with spin-then-park
waiting is suitable for use as the inner lock.  The thread that manages to acquire the 
inner lock is called the \emph{standby} thread -- there is at most one standby thread 
per lock at any given moment.  The inner lock constitutes a so-called slow path.  
The standby thread then proceeds to contend for the outer lock.  Again, it uses 
a spin-then-park waiting policy.  When the standby thread ultimately acquires the outer lock
it can enter the critical section.   At unlock time, if the current owner acquired
the lock via the slow path, it releases both the outer lock and the inner lock.  Otherwise
if it releases the outer lock and, if a standby thread exists, it unparks that standby
thread as the heir presumptive \footnote{This design admits a minor but 
useful optimization for single-threaded latency.  Normally the store to
release the lock would need to be followed by a store-load memory barrier (fence)  before the load
that checks for existence and identity of the standby thread.  That barrier can be 
safely elided if the standby thread uses a timed park operation that returns periodically.
Instead of avoiding the race -- which arises from architectural reordering and could
result in progress failure because a thread in unlock fails to notice a just-arrived waiting
thread -- we tolerate the race but recover as necessary via
periodic polling in the standby thread.  On platforms with expensive barrier operations,
this optimization can improve performance under no or moderate contention.
At any given time, at most one thread per lock -- the standby thread -- will be using
a timed park.  Creating and destroying the timers that underly a timed park
operation is not scalable on many operating systems, so constraining timer-related
activity is helpful.  In general, locks that allow parking typically require an atomic
or memory fence in the unlock() uncontended fast path, but this optimization avoids that expense,
while still avoiding stranded threads and missed wakeups.
}. 

The ACS consists of the owner, threads passing through their non-critical sections, and threads spinning
in the fast path arrival phase.  The PS consists of threads waiting for the inner
lock.  The standby thread is on the cusp and is transitional between the two sets. 
Under steady state the system converges to a mode where we have a stable
set of threads circulating over the outer lock (the ACS), at most one
thread spinning or parking in the standby position, and the remainder 
of the threads are blocked on the inner locks (the PS).  

We impose long-term fairness by detecting that the standby thread has
waiting too long and is ``impatient'', in which case it
requests direct handoff of ownership to the standby thread upon the next unlock operation. 
This construction attempts to retain the desirable properties of TAS-based
lock while providing CR and long-term fairness.  The result is a hybrid that
uses competitive handoff in most cases, reverting to direct handoff as part 
of an anti-starvation mechanism when the standby thread languishes too long. 

\Invisible{
*  CR makes choice of ``main'' outer lock less important -- reduces 
   sensitivity of performance to ``main'' lock.
*  Augmentation
*  Properties imposed by inner lock : CR; NUMA-awareness; fairness
*  General transformation like cohort locks;  wrap abstract outer
   lock with CR ``throttling'' construct.  Throttling provides K-exclusion.
*  patience: count of bypasses; time; integral of all waiters; maximum of all waiters
} 

%% The following paragraph is optional.  Exclude as needed for space ...
%% OPTIONAL
Ideally, we'd prefer to constrain the flow of threads from the PS into the ACS. 
A simple expedient is to make standby thread less aggressive than
arriving threads when it attempts to acquire the outer lock.  A related optimization
is to briefly defer waking the standby thread in the unlock path.  If the lock is
acquired by some other thread in the interim, then there is no need to unpark
the standby thread. In a similar fashion, the Solaris \texttt{pthread\_mutex} implementation attempts
to defer and hopefully avoid the need to awake potential successors.  
This defer-and-avoid strategy tends to keep the ACS stable and also avoids unpark latencies. 
Finally, another simple optimization is to constrain the number of threads allowed
to spin on the outer lock in the arrival phase.  Similarly, the Solaris
\texttt{pthread\_mutex} implementation bounds the number of threads concurrently
spinning on a lock 
%% OPTIONAL footnote
%% Consider : separate appendix for Solaris implementation ...
\footnote{For reference, the Solaris \texttt{pthread\_mutex} implementation
uses a simple polite test-and-test-and-set lock with a bounded spin duration.  
The TAS lock admits indefinite bypass and unfairness. 
If the spin
attempt fails, the thread reverts to a slow path where it enqueues itself and parks --
classic spin-then-park waiting.  The ``queue'' is mostly-LIFO \cite{OpenSolarisMutex}
and thread priorities are ignored. 
\Exclude{An inner lock protects the queue.} 
The implementation also bounds the number of concurrent spinning threads and uses 
the \texttt{schedctl} facility to avoid spinning if the owner is not iself running on a CPU.  
After releasing the lock in \texttt{pthread\_mutex\_unlock}, the implementation 
checks if the queue is empty.  If so, it returns immediately.  Otherwise it waits 
briefly to see if the lock happens to be acquired in the interim by some other thread.  If so, 
the caller can return without needing to dequeue and unpark an heir presumptive.  
The responsibility for succession and progress is delegated to the new owner.  
%% Wording: passes to; is delegate to
Such \textit{unpark avoidance} reduces the voluntary context switch park-unpark 
rate and reduces the latency of the unlock operator.  
This defer-and-avoid strategy also tends to keep the ACS stable.
The policies of bounding the number of concurrent spinners and 
unpark avoidance act toward constraining the size of the ACS.
%% wording: act to effective; act to; act toward; tend to; 

The implementation also provides \textit{wait morphing} -- if a 
\texttt{pthread\_cond\_signal} operation selects a thread that waits on a mutex 
held by the caller, then that thread is simply transferred from the condition 
variable's wait-set directly to the mutex's queue of blocked threads, avoiding 
the need to unpark the notifyee.   This operation is fast, and reduces the hold 
time when \texttt{pthread\_cond\_signal} is called within a critical section.  
In addition, we avoid waking a thread while the lock that thread is trying to 
acquire is already held by the caller, reducing futile and unnecessary
contention between the notifier and notifyee.  Morphing leverages the
observation that is is usually safe to shift a \texttt{pthread\_cond\_signal} call
from within a critical section to immediately after the critical section.
}. 

\Invisible{Delegate; pass; shirk; transfer; convey; impart;} 

Arriving threads start with global spinning on the outer lock, and if
they can't manage to obtain the lock within the arrival spinning phase, they then 
revert to the MCS lock, which uses local waiting.  Global spinning allows
more efficient lock hand-over, but local spinning generates less coherence
traffic and provides gracefully performance under high contention \cite{asplos94-lim}. 
Threads waiting on the inner
MCS lock simply spin or spin-then-park on the thread-local variable,
avoiding concerns about back-off policies. 
All park-unpark activity takes place on paths outside the critical section.
The inner lock provides succession by direct handoff via MCS,
while the outer lock provides succession by competitive handoff.  
This constitutes a 3-stage waiting policy : threads first spin
globally; then, if necessary, enqueue and spin locally; and then park.   

%% OPTIONAL ...
The LOITER  transformation allows us to convert a lock such as MCS,
which uses direct handoff, into a composite form that allows a fast path
with barging.  The resultant composite LOITER lock enjoys the benefits of both 
direct handoff and competitive succession, while mitigating the undesirable 
aspects of each of those policies. 
Specifically, the new construct uses direct handoff for threads 
in the slow contention path, but allows competitive succession for threads
circulating outside the slow path,
retaining the best properties of both MCS and TAS locks.  

%% OPTIONAL ...
\Invisible{
To further restrict and constrain concurrency, the implementation 
can restrict or cap the number of threads spinning on a lock at any given moment.} 

%% OPTIONAL ...
A useful complementary thread-local policy in the spinning phase implementation is 
to abandon the current spin episode if the TAS  atomic operation on
the outer lock fails too frequently.  This condition indicates a sufficient flow of
threads in the ACS over the lock.  \Exclude{An implementation might use a simple
thread-local and episode-local
count of the number of such failures and abandon the spin attempt when
the count exceeds some bound.  Equivalently, after a TAS failure
we might use a Bernoulli trial -- via simple uniform PRNG with thread-local
state -- to decide whether to abandon the spin attempt.}    
Another variation is to monitor either traffic over the lock
or the arrival or spinners, and to abandon the spin attempt if the rate
or flux is too high.  By abandoning the spin attempt early, the thread
reverts from spinning to parking.  This is tantamount to self-culling. 

%% OPTIONAL ...
If the inner lock is NUMA-friendly -- say, a cohort lock -- then the
aggregate LOITER lock is NUMA-friendly. As threads circulate between
the active and passive sets, the inner lock tends to filter out threads
from different nodes, and the ACS then tends to converge toward a set
of threads located on a given node.  Decreased NUMA-diversity of the 
ACS decreases lock migration rates and yields better throughput. 
   
\subsection{LIFO-CR} 

This design starts with a pure LIFO lock 
\footnote{If we use a pure LIFO lock then the LWSS should correspond to 
the ACS size, giving an easy way to measure the ideally minimal ACS size and maximum
benefit afforded by CR.} 
with an explicit stack of waiting threads.  Contended threads push an MCS-like node onto
the stack and then spin or spin-then-park on a thread-local flag.  
These nodes can be allocated on stack.  
When threads are
waiting, the unlock operator pops the head of stack -- the most recently arrived
thread -- and directly passes ownership to that thread.  
(We also define a special distinguished value for the stack pointer that indicates
the lock is held and there are no waiters.  0 indicates that the lock is not held). 
Both ``push'' and ``pop'' operations are implemented via atomic compare-and-swap CAS
instructions.
Only the lock holder can ``pop'' elements, so the approach is immune to ABA pathologies.  
The stack is multiple-producer but, by virtue of the lock itself, single-consumer.  
The ACS consists of the owner, the threads circulating through their respective
NCS regions, and the top of the stack.  The PS consists of the threads deeper
on the stack.  Admission order is effectively cyclic round-robin
over the members of the ACS, regardless of the prevailing LIFO lock admission policy.
We then augment the lock to periodically pick the tail of the 
stack -- the eldest thread -- to be the next owner.  
This imposes long-term fairness  
We refer to the resultant lock as \textbf{LIFO-CR}.
LIFO admission order may improve temporal locality and reduce misses in shared caches.  
Both LIFO-CR and LOITER offer performance competitive with MCSCR.  

\Invisible{LIFO;  MRA most-recently-arrived is warmest and thus fastest; Fastest Thread next admitted;} 

%% OPTIONAL 
It is relatively simple to augment any given unfair lock so that starving threads 
are periodically given a turn via direct handoff.   The Solaris and windows schedulers 
employ similar anti-starvation policies.  If threads languish too long on the run 
queue because their effective priority is too low, then they'll be given transient 
priority boosts until they run.   By analogy, this policy can extend to locks, where
waiting threads that languish too long can be explicitly granted ownership.  
This allows our locks to enjoy the benefits of short-term unfairness but
explicitly manage long-term unfairness and to ensure eventual progress.

%% OPTIONAL ...
Normally the ``pop'' operator would employ CAS in a loop.  We can avoid the
loop and, as an optional optimization, implement a constant-time unlock operation 
by observing that if the CAS fails, 
then new threads have arrived and pushed themselves onto the stack, and there are at
least two elements on the stack.  We can thus implement a plausibly LIFO unlock by 
naively unlinking and passing ownership to the element that follows the thread 
identified by the failed CAS return value.

%% OPTIONAL ...
Under LIFO-CR both arriving and departing (unlocking) threads will update the 
the head of the stack, potentially creating an undesirable coherence hot-spot.  
MCSCR avoids this concern.  In practice, however, this does not seem to adversely 
impact performance in LIFO-CR.  The performance properties of the inner-outer lock and LIFO-CR 
are approximately the same as MCSCR.  This algorithm works particularly well
with spin-then-park waiting policies, as the threads near the top of the stack
are most likely to run next, but are also the most likely to be spinning instead
of blocked, thus avoiding expensive context switching and unpark activity. 

\Invisible{
*  Stack prefix; suffix
*  epicycles
} 

\Invisible{
CR provides a number of modes of benefit.   I'll start with the cache 
occupancy/residency argument you noted above.   Lets assume our CS and NCS 
exhibit reasonable temporal locality and reuse patterns.   Critically, 
the NCS operations can erode the cache residency of the data accessed in the CS.   
(I am assuming a single-socket environment with a large share last-level cache.   
All the interesting cache effects happen in the LLC).   The more threads 
circulating, the greater the decay rate of the CS data from the cache.    
If you know the LLC capacity and can control the CS and NCS ``length'' and 
``width'' (footprint -- amount of data accessed), it's then pretty easy to set up a 
contrived case to show the effect.   Note that a CS ``activation doesn't have to 
be long or touch much data to see the impact.   A nice example I'm using is 
where the CS looks up a randomly selected key in a red-black tree.  Any 
individual CS is relatively short and does not touch much data, but the key is 
the size of the tree.   If we restrict concurrency, then the CS suffers fewer 
misses arising from displacement caused by NCS accesses.   The CS then runs 
faster, and we enjoy better throughput.   
}

\section{Scheduler Interactions} 

%% Relegate-Demote this to appendix
%% OPTIONAL ...

%% Background material ...
%% OPTIONAL ...
The operating system kernel scheduler provides 3 states for threads : \emph{running}, 
\emph{ready}, and \emph{blocked}.  Running indicates the thread is active on a processor.
Ready indicates the thread is eligible to run, but has not been dispatched onto
a processor.  Blocked indicates the thread is suspended and ineligible for 
dispatch and execution -- the thread is typically waiting for some condition.  
The park operator transitions a running thread to blocked state and 
unpark makes a blocked thread ready.  The kernel typically manages all 
ready-running transitions while the lock subsystem, via park-unpark, controls 
the ready-blocked transitions associated with locking.  The kernel scheduler's dispatch function
shifts a thread from ready to running.  Involuntary preemption via time-slicing
shifts a thread from running to ready.  Intuitively, park causes a thread to 
sleep and unpark wakes or resumes that thread, reenabling the thread for 
subsequent dispatch onto a processor.   A parked thread is waiting for
some event to occur and notification of that event occurs via a corresponding 
unpark.  We expect the scheduler is itself work conserving with
respect to idle CPUs and ready threads.  In addition, ready threads will
eventually be dispatched and make progress.  
%% OPTIONAL
If there are available idle CPUs, unpark($T$) will dispatch
T onto one of those CPUs, directly transitioning $T$ from blocked to running.
If there are more ready threads than CPUs then the kernel will use preemption
to multiplex those threads over the set of CPUs.  Threads that are ready
but not running wait for a time slice on dispatch queues. 
\Exclude{An ideal \texttt{sched\_yield} implementation would first determine 
if there are any other ready threads.  If not, it returns immediately.  
Otherwise it transitions the caller from running to ready and picks and
dispatches one of those other ready threads, allowing it to run.} 
\Invisible{A \emph{thread} is simply a software construct exposing
a virtualized logical processor.} 

%% OPTIONAL ...
Preemption is controlled by the kernel and reflects an involuntary context switch.   
The victim is changed from running to ready and some other ready thread is dispatched
on the CPU and made to run.  Preemption is usually triggered by timer interrupts. 
Typically the kernel resorts to preemption when there are more runnable threads 
than CPUs.  The kernel preempts one thread $T$ running on CPU $C$ in order to allow 
some other ready thread a chance to run on $C$.  Preemption provides long-term 
fairness over the set of runnable threads competing for the CPUs.  
The kernel uses preemption to multiplex $M$ ready threads over $N$ CPUs, where $M > N$. 
When a thread is dispatched onto a CPU  it receives a time slice (quantum).  When the 
quantum expires, the thread is preempted in favor of some ready thread. 
Threads that have been preempted transition from running to ready state.  

%% OPTIONAL: ...
A CPU is either \emph{idle} or \emph{running}.
A CPU becomes idle when the operating system has no ready threads to
dispatch onto that CPU.  When a thread on a CPU parks itself and the 
operating system (OS) scheduler is unable to locate another suitable ready 
thread to dispatch onto that CPU, the CPU transitions to idle.  
Subsequently, an idle CPU $C$ switches back to running when some blocked thread $T$ 
is made ready via unpark and the OS dispatches $T$ thread onto $C$.
%% transitioning $T$ from ready to running and transitioning CPU $C$ from idle to running.  
Thread park-unpark transitions can induce CPU running-idle transitions. 

%% OPTIONAL ...
The longer a CPU remains idle, the deeper the reachable sleep state.  
Deeper idle (sleep) states draw less power, and
allow more aggressive turbo mode for sibling cores, but such sleep states
take longer to enter and exit.  Specifically, to leverage the benefits
of deeper sleep states, the CPU needs to stay in that state for some
period to amortize the entry and exit costs.  As such, we prefer to avoid 
frequent transitions between idle and running states for CPUs.
When a thread on a CPU parks and the operating system (OS) scheduler is unable to locate
another suitable ready thread to dispatch onto that CPU, the CPU
becomes idle.  Subsequently, an idle CPU $C$ switches to running when 
some blocked thread $T$ is made ready via unpark and the OS dispatches
$T$ thread onto $C$, transitioning $T$ from ready to 
running and transitioning CPU $C$ from idle to running.  
Thread park-unpark transitions can induce CPU idle transitions. 

%% OPTIONAL ...
Frequent park-unpark activity may cause rapid transitions between
idle and running CPU states, incurring latencies when unpark dispatches a thread
onto a idle CPU and that CPU exits idle state.  Furthermore, frequent 
transitions in and out of idle may prevent a CPU from reaching deeper power saving
idle (sleep) states \footnote{the CPU may not remain idle sufficiently long to
reach those deep sleep states.}.   

%% via \usepackage{listings}
%% make sure to expand tabs : in vim : set ts=4; expandtab; retab!
%% also, clean up mac UTF characters
%% \rowcolors{2}{gray}{white}
%% \lstset{language=c,frame=single,numbers=left,columns=fullflexible} 
%% \lstset{basicstyle=\scriptsize,framesep=5pt}
%% \lstset{xleftmargin=20pt,caption=Partitioned Ticket-Lock Algorithm,label=Algorithm} 
%% \lstinputlisting{Algorithm.c}

\end{document}